\documentclass[final]{siamltex}
\usepackage{amsfonts,amsmath,mathrsfs,bm}
\usepackage{amssymb}
\usepackage[notref,notcite]{showkeys}
\usepackage{hyperref}
\usepackage{graphicx}
\usepackage{wrapfig}
\usepackage{subfig}
\usepackage{float}
\usepackage{bbm}
\usepackage{algorithmic}
\usepackage{gensymb,color}
\usepackage{tikz}
\usepackage{tikz-3dplot}

\newcommand{\N}{\mathbb{N}}
\newcommand{\R}{\mathbb{R}}

\newcommand{\E}{\mathbb{E}}

\newtheorem{remark}[theorem]{Remark}
\newtheorem{algorithm}{Algorithm}

\DeclareMathOperator{\argmin}{\rm arg\, min}
\DeclareMathOperator{\argmax}{\rm arg\, max}

\title{Edge-promoting adaptive Bayesian experimental design for X-ray imaging}

\author{
T. Helin\footnotemark[2]
\and N. Hyv\"onen\footnotemark[3]
\and J.-P. Puska\footnotemark[3]
}

\begin{document}
\maketitle

\renewcommand{\thefootnote}{\fnsymbol{footnote}}
\footnotetext[2]{LUT University, School of Engineering Science, P.O.~Box 20, FI-53851 Lappeenranta, Finland. The work of TH was supported by the the Academy of Finland (decisions 320082 and 326961).}
\footnotetext[3]{Aalto University, Department of Mathematics and Systems Analysis, P.O.~Box 11100, FI-00076 Aalto, Finland (nuutti.hyvonen@aalto.fi, juha-pekka.puska@aalto.fi). The work of NH and JP was supported by the Academy of Finland (decision 312124).}

\begin{abstract}
  This work considers sequential edge-promoting Bayesian experimental design for (discretized) linear inverse problems, exemplified by X-ray tomography. The process of computing a total variation type reconstruction of the absorption inside the imaged body via lagged diffusivity iteration is interpreted in the Bayesian framework. Assuming a Gaussian additive noise model, this leads to an approximate Gaussian posterior with a covariance structure that contains information on the location of edges in the posterior mean. The next projection geometry is then chosen through A-optimal Bayesian design, which corresponds to minimizing the trace of the updated posterior covariance matrix that accounts for the new projection. Two and three-dimensional numerical examples based on simulated data demonstrate the functionality of the introduced approach.
\end{abstract}

\renewcommand{\thefootnote}{\arabic{footnote}}



\pagestyle{myheadings}
\thispagestyle{plain}
\markboth{T. HELIN, N. HYV\"ONEN, AND J.-P. PUSKA}{ADAPTIVE EXPERIMENTAL DESIGN IN X-RAY IMAGING}

\section{Introduction}
\label{sec:introduction}

Large-scale Bayesian inverse problems have rapidly gained popularity during the last two decades \cite{Kaipio06, stuart_2010}. While computational resources seem ever-increasing, data acquisition in a number of real-life inverse problems remains restricted or expensive. In consequence, there is a growing interest to develop computational methodologies for designing efficient data acquisition techniques or experimental setups to maximize the value of data in the solution process. Bayesian {\em optimal experimental design} (OED) provides a principled approach to such a task, and it has been widely adopted in the inverse problems community; see,~e.g.,~\cite{alexanderian2021optimal_review} and reference therein.

A Bayesian optimal design $p^*$ maximizes the expected utility function $U(p)$ over the design space $D$ with respect to the data $y$ and model parameters $u$ according to
\begin{eqnarray}
	\label{eq:OED_task}
	p^* & = & \underset{p\in D}{\argmax} \, \E[U(p; u,y)] \nonumber \\
	& = & \underset{p\in D}{\argmax} \int_Y \int_\Theta U(p; u, y) \pi(u \, | \, p, y) \pi(y \, | \, p) \, d u \, dy.
\end{eqnarray}
Here $\pi(u \, | \, p, y)$ and $\pi(y \, | \, p)$ stand for the posterior distribution of the parameter $u$ and the marginal distribution of the data $y$, respectively, under the design $p$. The utility function can be devised in a number of ways; the two most common choices for $U$ are arguably a \emph{negative quadratic loss function} that measures the squared distance from $u$ to a specific point estimator such as the posterior mean and the \emph{expected information gain} where $U$ is the Kullback--Leibler distance between the posterior and prior distributions.

The computational crux of \eqref{eq:OED_task} lies with the double integral over the potentially high dimensional parameter and data spaces related to the considered imaging application. Moreover, if the set of possible designs is vast (e.g.,~$p$ is a continuous parameter on a high-dimensional manifold), an exhaustive search may seem unfeasible. Still, significant progress has been made in the past working under conditions that allow closed form presentations for the above double integral. For the aforementioned two cost functions, the integrals in \eqref{eq:OED_task} are explicitly solvable when the posterior and data marginal distributions are Gaussian. In inverse problems, this occurs when the forward operator is linear, and the prior and additive noise distributions are Gaussian~\cite{Kaipio06}. In such a case, the double integral is proportional to the
the trace and the determinant of the posterior covariance, respectively.
In the literature, these are called the Bayesian A and D-optimality criteria~\cite{chaloner1995bayesian}.

There has also been substantial effort to go beyond the conditions that enable explicit integration in~\eqref{eq:OED_task}. In this regard, important early work was carried out in \cite{ryan2003estimating, huan2010accelerated, huan2013simulation} toward developing fast double loop Monte Carlo algorithms for tackling general inverse problems. More recent approaches have concentrated on improving efficiency of integral approximations by Laplace's method in the context of nonlinear inverse problems \cite{long2013fast, alexanderian2016fast, crestel2017optimal, beck2018fast}. Under well-designed approximation schemes, the computational complexity of such methods can be low in terms of the number of required forward solutions and scalable in the sense of being independent of the parameter and data dimensions \cite{wu2020fast, wu2021fast}.

This paper grows out of the observation that the efficient use of non-Gaussian prior distributions in Bayesian OED for inverse problems has not been addressed in the literature. Indeed, successful solvers in imaging problems rely on well-designed prior information, which in variational regularization is often formulated in terms of nonquadratic penalty functionals \cite{scherzer2009variational}.
Following the popularity of convex regularization in imaging, similar ideas have been successfully introduced to the Bayesian setting by formulating non-Gaussian priors in Banach spaces such as Besov spaces or BV spaces; see,~e.g.,~\cite{wang2017bayesian, yao2016tv, vanska2009statistical, Lassas2009, agapiou2018sparsity, agapiou2018rates, lv2020nonlocal}. Motivated by these observations, our work contributes toward including non-Gaussian prior distributions in Bayesian OED practices for inverse problems and imaging.

\subsection{Our contribution}
This work introduces a computational method for performing greedy sequential OED for linear inverse problems with a {\em total variation} (TV) prior. The proposed algorithm is novel, as it does not utilize Laplace's approximation or sampling schemes to tackle a non-Gaussian posterior distribution. Instead, its founding idea is based on the so-called lagged diffusivity approximation for TV introduced in \cite{Vogel96}. At each step of the sequential algorithm, a lagged diffusivity iteration is employed to produce a sequence of Gaussian approximations for the TV prior, presumably with increasing accuracy close to the posterior mode. Assuming an additive Gaussian noise model and one of the two cost functions $U$ considered above, replacing the TV prior by its final approximation allows a closed form solution for the double integral in \eqref{eq:OED_task}. This leads to a standard form A or D-optimality criterion for choosing the (subsequent) measurement design.

Like the lagged diffusivity approximation, our method could also be formulated for a large class of Gibbs prior measures. Moreover, it may be possible to extend some of our ideas to the framework of nonlinear inverse problems by combining them with Laplace's method. Be that as it may, in this work the proposed algorithm is only tested with a linear inverse problem and a TV prior.

We develop our method in the context of X-ray tomography and A-optimality, building upon our previous work \cite{Burger21} that considered efficiency and adaptivity of sequential OED in such a framework. X-ray tomography is particularly well-suited for a sequential approach to OED as the radiation exposure (i.e.~the number of projections) often needs to be minimized while maximizing the quality of the reconstruction in certain regions of interest, the locations of which may be unknown {\em a priori}.

We consider X-ray tomography in both two and three-dimensional imaging setups with a narrow X-ray beam whose propagation angle and lateral position can be optimized. Our main hypothesis is that an (approximate) TV prior in Bayesian OED for X-ray tomography should promote designs that efficiently recover edges in the imaged target. The presented numerical experiments, which are based on simulated data, demonstrate that our algorithm does indeed perform well for certain piecewise constant phantoms when compared with the use of equiangular full-width projections corresponding to an equivalent radiation dose.

This text is organized as follows. Section~\ref{sec:mm} introduces a discretized linear measurement model for X-ray tomography. In Section~\ref{sec:lagged_diff} the basic ideas of lagged diffusivity iteration are interpreted in the Bayesian framework. The concept of A-optimality is recalled in Section~\ref{sec:A_optimality}, and it is subsequently combined with the lagged diffusivity iteration to form a sequential OED algorithm in Section~\ref{sec:opti}. The numerical experiments are presented in Section~\ref{sec:numerics}, and the concluding remarks are listed in Section~\ref{sec:conclusion}.

\subsection{Literature review}
Bayesian OED has gained substantial attention in large-scale inverse problems during the recent years. In addition to the works mentioned above, let us list \cite{alexanderian2018efficient, attia2018goal, fohring2016adaptive, haber2008numerical, haber2009numerical, haber2012numerical, hyvonen2014eit, khodja2010guided,long2015fast,long2013fast, aretz2020sequential} to name a few papers on this topic. In particular, there is an interesting line of research developing Bayesian OED for infinite-dimensional inverse problems \cite{alexanderian2014optimal, alexanderian2016bayesian, alexanderian2016fast, alexanderian2021optimal}. Here, we test our novel ideas in a sequential optimization strategy, which has previously been formalized for large-scale problems in \cite{huan2016sequential} based on ideas from dynamical programming. For general references on Bayesian OED, we mention the review papers \cite{chaloner1995bayesian, ryan2016review} and the monograph \cite{pukelsheim2006optimal}.

Optimization of the imaging geometry in X-ray tomography has previously been considered in \cite{ruthotto2017optimal,Burger21}. The former article explored empirical A-optimal design in constrained problems based on training data by adopting sparse sensor-placing strategies and a gradient-based optimization scheme.  The latter paper \cite{Burger21} introduced more degrees of freedom (lateral position of the source-receiver pair) to the problem, considered efficient evaluation of the A and D-optimality target functions and introduced adaptivity to the algorithm.

The idea of TV denoising was originally presented in~\cite{Rudin92}, and the lagged-diffusivity fixed point iteration for approximating TV regularization was introduced in~\cite{Vogel96}. The convergence of the algorithm has been considered,~e.g.,~in \cite{dobson1997convergence,chan1999convergence} for finite-dimensional image restoration problems. Finally, let us remark that total variation regularization is widely employed in computed tomography; see,~e.g.,~\cite{sidky2008image,liu2012adaptive,tian2011low}.

\section{Measurement model and its discretization}
\label{sec:mm}
The X-ray measurements are modeled either as parallel beam or cone beam tomography, where multiple rays are directed into the object $D \subset \R^d$, $d=2$ or $3$, and the resulting intensities of the rays are measured at detectors~\cite{Natterer01}. The attenuation is described by the equation
\begin{equation}
	\label{eq:model}
	I = I_0 \exp \left( - \int_{L} u \, {\rm d}s \right),
\end{equation}
where $L$ is the line along which the considered ray travels, $I_0$ is the intensity of the X-ray before entering the object and $u: D \to \R_+$ is the absorption. Obviously, \eqref{eq:model} can equivalently be given as
$$
\log(I_0) - \log(I) = \int_{L} u \, {\rm d} s.
$$
In particular, the difference between the logarithms of the  emitted and measured intensities is typically considered as the available data when X-ray tomography is tackled mathematically.

We discretize the imaged domain into $n' \in \N$ pixels or voxels, but assume the absorption distribution vanishes at the boundary pixels/voxels and denote the  number of interior pixels/voxels by $n < n'$. The forward operator, mapping the discretized absorption to a {\em single} set of log-intensity measurements at the detectors, can be approximated by a matrix $R \in \mathbb{R}^{m \times n}$, where $m$ is the  number of detectors (see,~e.g.,~\cite{Siltanen03}); typically the dimension of the unknown is higher than the number of pixels in a {\em single} projection image, i.e.~$m\ll n$. In what follows, we abuse the notation by denoting with $u \in \R^n$, $n \in \N$, both the vector of pixel/voxel values defining the discretized (interior) absorption as well as a (smooth enough) function on $D$ taking the given absorption values at the center points of the respective pixels/voxels. The correct interpretation should be clear from the context.

\section{Total variation prior and lagged diffusivity}
\label{sec:lagged_diff}
Let $u_{k-1} \in \R^n$ be the reconstruction after taking $k-1 \in \N_0$ X-ray projections and assume that the $k$th projection image has just become available; Section \ref{sec:A_optimality} below explains how the experimental design for this newest projection was chosen. Let us denote by
$$
\mathbf{R}_k =
\begin{bmatrix}
  R(p_1) \\
  \vdots \\
  R(p_k)
\end{bmatrix} \in \R^{km \times n} \qquad {\rm and} \qquad
\mathbf{y}_k = \begin{bmatrix}
  y_1 \\
  \vdots \\
  y_k
\end{bmatrix} \in \R^{km}
$$
the stacked X-ray matrix corresponding to all previous projections and the corresponding stacked noisy data vector, respectively. The vectors $p_1, \dots, p_k$ are the design parameters employed thus far. The measurements $y_1, \dots, y_k$ are modeled as realizations of the random variables
\begin{equation}
  \label{eq:meas_model}
Y_j = R(p_j) U + N_j, \qquad j=1, \dots, k,
\end{equation}
where $U$ is the randomized discrete absorption and the noise $N_j$ is assumed to follow a zero-mean Gaussian distribution $\mathcal{N}(0, \Gamma_{\rm noise}^{(j)})$, where is $\Gamma_{\rm noise}^{(j)} \in \R^{m \times m}$ is symmetric and positive definite. The noise processes $N_1, \dots, N_k$ are assumed to be mutually independent.

The (accurate) prior for the absorption $U$ has an edge-promoting probability density of the form
\begin{equation}
\label{eq:sigma_prior}
\pi(u) \propto \exp \! \big( -\gamma  \Phi(u) \big),
\end{equation}
where $\gamma > 0$ is a free parameter and $\Phi$ is defined through
\begin{equation}
\label{eq:aRRa}
\Phi(u) = \int_{D} \varphi \big(|\nabla u | \big) \, {\rm d} x,
\end{equation}
accompanied by the information that $u$ vanishes at the pixels/voxels next to the boundary of $D$. In this work, we exclusively consider the (smoothened) TV prior~\cite{Rudin92}
\begin{equation}
\label{eq:arra}
\varphi(t) = \sqrt{t^2 + T^2} \approx | t |,
\end{equation}
where $T>0$ is a small parameter that ensures differentiability. However, it would also be possible to consider other edge-preferring priors such as Perona--Malik~\cite{Perona90}.

According to the Bayes' formula and assuming the measurement model \eqref{eq:meas_model}, the posterior density for $u$ thus reads
\begin{align}
  \label{eq:post_u}
\pi(u \, | \, \mathbf{y}_k) \, &\propto \, \pi( \mathbf{y}_k \, | \, u) \, \pi(u)  \nonumber \\
&\propto \, \exp \Big(-\frac{1}{2}  (\mathbf{y}_k - \mathbf{R}_k u )^{\rm T} (\bm{\Gamma}_{\rm noise}^{(k)})^{-1}(\mathbf{y}_k - \mathbf{R}_k u ) - \gamma \Phi(u) \Big),
\end{align}
where $\bm{\Gamma}_{\rm noise}^{(k)} := {\rm diag}(\Gamma_{\rm noise}^{(1)}, \dots, \Gamma_{\rm noise}^{(k)})\in \R^{km \times km}$ is a block diagonal matrix defined by the noise covariance matrices for the previous measurements.
Our leading idea is to iteratively approximate $\Phi(u)$ by quadratic terms in the spirit of the lagged diffusivity iteration~\cite{Vogel96}; see also~\cite{Arridge13,Harhanen15}. This results in an iterative algorithm for computing the reconstruction $u_k$ after $k$ measurements as well as forming the corresponding covariance matrix employed in choosing the next projection geometry by means of A-optimality.

\subsection{First step: Gaussian approximation for the prior around $u_{k-1}$}
Let $\{ \phi_j\}_{j=1}^{n'} \subset H^1(D)$ be a Lagrangian finite element basis for the dual mesh of the employed pixelification/voxelification for $D$ numbered so that the first $n$ basis functions correspond to the interior pixels/voxels in $D$. In particular, the $j$th basis function $\phi_j$ takes value one at the midpoint of the $j$th pixel/voxel and vanishes at all the other midpoints. After identifying $u$ with its interpolant in this basis and recalling that $u$ is assumed to vanish at (the midpoints of) the boundary voxels, one easily deduces that
$$
\nabla_{\!u} \Phi(u) = H(u)u, \qquad u \in \R^n,
$$
where
\begin{align}
\label{eq:H}
H_{i,j}(w)
&:= \int_{D} \frac{1}{\sqrt{|\nabla_{\!x} w(x)|^2 + T^2}} \, \nabla \phi_i(x) \cdot \nabla \phi_j(x) \, {\rm d} x,
\qquad i,j=1,\dots, n,
\end{align}
for any $w \in \R^n$ interpreted as an element of $H^1(D)$ via the introduced finite element basis.

Observe that $H(w) \in \R^{n \times n}$ is the stiffness matrix for a finite element approximation of the differential operator
\begin{equation}
\label{eq:diffop}
- \nabla \cdot \big( \rho (|\nabla w|)  \nabla (\, \cdot \,) \big)
\end{equation}
over $D$, with
$$
\rho(v) := \frac{1}{\sqrt{v^2 + T^2}}
$$
and a homogeneous Dirichlet condition on $\partial D$. As a consequence, $H(w)$ is positive definite and, in particular, invertible for any $w \in \R^n$.

Let us then introduce the quadratic penalty function
$$
  \Phi_{k-1}(u) = \frac{1}{2} u^{\rm T} H(u_{k-1}) u + \frac{1}{2} u_{k-1}^{\rm T} H(u_{k-1})  u_{k-1} + \int_{D} \frac{T^2}{\sqrt{|\nabla_{\! x} u_{k-1}(x)|^2 + T^2}} \, {\rm d} x.
  $$
It is straightforward to check that
  \begin{equation}
\label{eq:approx}
\Phi_{k-1}(u_{k-1}) = \Phi(u_{k-1}) \quad {\rm and} \quad
\nabla_{\!u} \Phi_{k-1}(u_{k-1}) = \nabla_{\!u} \Phi(u_{k-1}) = H(u_{k-1}) u_{k-1},
\end{equation}
  meaning that the tangent planes
  for the graphs of $\Phi_{k-1}: \R^n \to \R_+$ and $\Phi:  \R^n \to \R_+$ coincide above the previous reconstruction $u_{k-1}$. Substituting $\Phi$ for $\Phi_{k-1}$ in \eqref{eq:post_u}, we have thus arrived at the approximate Gaussian posterior density
\begin{equation}
  \label{eq:approx_post}
  \pi^{(1)}(u \, | \, \mathbf{y}_k) \propto \exp \Big(-\frac{1}{2}  \big( (\mathbf{y}_k - \mathbf{R}_k u )^{\rm T} (\mathbf{\Gamma}_{\rm noise}^{(k)})^{-1}(\mathbf{y}_k - \mathbf{R}_k u ) + \gamma u^{\rm T} (\Gamma_{k-1}^{(1)})^{-1} u \big) \Big),
\end{equation}
where $\Gamma_{k-1}^{(1)} :=  H(u_{k-1})^{-1}$.

\subsection{Second step: iterating the argument}
Building the initial Gaussian approximation \eqref{eq:approx_post} for the posterior \eqref{eq:post_u} consists essentially of two steps: (i) assuming a reasonable estimate $u_{k-1}$ for the solution of the studied inverse problem and (ii)~forming the approximate prior covariance via $\Gamma_{k-1}^{(1)} =  H(u_{k-1})^{-1}$. Introducing the mean of the density \eqref{eq:approx_post} as a new, hopefully more accurate reconstruction and iterating the argument leads to a Bayesian interpretation of the lagged diffusivity algorithm~\cite{Vogel96} for computing a reconstruction $u_k$ after having $k$ projection images in hand:

Define $u_{k-1}^{(0)} = u_{k-1}$. Assuming the availability of $u_{k-1}^{(j-1)}$, form an approximate prior covariance
\begin{equation}
  \label{eq:step1}
\Gamma_{k-1}^{(j)} =  H(u_{k-1}^{(j-1)})^{-1}.
\end{equation}
Introduce the corresponding posterior density
\begin{equation}
  \label{eq:approx_post2}
\pi^{(j)}(u \, | \, \mathbf{y}_k) \propto \Big(-\frac{1}{2}  \big( (\mathbf{y}_k - \mathbf{R}_k u )^{\rm T} (\mathbf{\Gamma}_{\rm noise}^{(k)})^{-1}(\mathbf{y}_k - \mathbf{R}_k u ) + \gamma u^{\rm T} (\Gamma_{k-1}^{(j)})^{-1} u \big) \Big)
\end{equation}
and compute its mean
\begin{equation}
  \label{eq:step2}
 u_{k-1}^{(j)} =  \Gamma_{k-1}^{(j)} \mathbf{R}_k^T \big(\mathbf{R}_k \Gamma_{k-1}^{(j)} \mathbf{R}_k^T + \gamma \bm{\Gamma}_{\rm noise}^{(k)}  \big)^{-1}\mathbf{y}_k;
\end{equation}
see.,~e.g.,~\cite{Kaipio06}.

If the chosen stopping criterion is satisfied at $j=J$, one dubs $u_k := u_{k-1}^{(J)}$ the reconstruction after $k$ projection images. The corresponding covariance matrix for the Gaussian density \eqref{eq:approx_post2} with $j=J$ is
\begin{equation}
    \label{eq:posterior_cov}
  \Gamma_{k} = \gamma^{-1} \big( \Gamma_{k-1}^{(J)} - \Gamma_{k-1}^{(J)} \mathbf{R}_k^T \big(\mathbf{R}_k \Gamma_{k-1}^{(J)} \mathbf{R}_k^T + \gamma \bm{\Gamma}_{\rm noise}^{(k)}  \big)^{-1} \mathbf{R}_k \Gamma_{k-1}^{(J)} \big);
\end{equation}
see,~e.g.,~\cite{Kaipio06}. This covariance structure is then used for choosing the parameter vector $p_{k+1}$ defining the next X-ray projection as explained in the following section.

\begin{remark}
  The two steps \eqref{eq:step1} and \eqref{eq:step2} correspond to a lagged diffusivity iteration for minimizing the argument of the exponential in \eqref{eq:post_u}, that is, computing an approximation of the {\em maximum a posteriori} (MAP) estimate for the absorption in $D$ after the availability of $k$  projection images. As the convergence of the lagged diffusivity iteration has been proven for denoising problems in~\cite{dobson1997convergence,chan1999convergence}, it is arguably not too far-fetched to hope that the above introduced iteration converges toward the mode of the posterior \eqref{eq:post_u}. For large enough $j$, the Gaussian density $\pi^{(j)}$ defined by \eqref{eq:approx_post2} can thus be considered an approximation for the exact posterior \eqref{eq:post_u} close to its mode, cf.~\eqref{eq:approx}.
\end{remark}

\section{A-optimal design}
\label{sec:A_optimality}

Let us assume that we have $k \in \N$ projection images of the imaged object $D$ at our disposal. According to the construction in the previous section, this leads to the (approximate, posterior) probability
distribution $\mathcal{N}(u_{k}, \Gamma_{k})$ for the absorption $U$, with the mean and covariance defined via \eqref{eq:step2} and \eqref{eq:posterior_cov}, respectively.
Assuming the new X-ray projection obeys the same measurement model as the previous ones, i.e.~\eqref{eq:meas_model}, the Gaussian posterior covariance after the $(k+1)$th projection reads
\begin{equation}
  \label{eq:p_postcov}
 \Gamma_{\rm post}^{(k+1)}(p) = \Gamma_{k} - \Gamma_{k} R(p)^T \big(R(p) \Gamma_{k} R(p)^T + \Gamma_{\rm noise}^{(k+1)}  \big)^{-1} R(p) \Gamma_{k},
\end{equation}
where $p$ is the to-be-selected design parameter determining the $(k+1)$th projection.

The task in hand is now to choose the $(k+1)$th projection, or more precisely, the corresponding design parameter $p_{k+1}$. In Bayesian optimal experimental design, one often considers minimizing the expected squared distance of the unknown in a given (semi)norm around the posterior mean; see,~e.g.,~\cite{alexanderian2016bayesian,chaloner1995bayesian} for more details. In the considered simple setting, this leads to the so-called A-optimality criterion for choosing the $(k+1)$th design parameter,
\begin{equation}
\label{eq:Aoptimal}
p_{k+1} =  \underset{p}{\argmin} \, {\rm tr}  \big(A \Gamma_{\rm post}^{(k+1)}(p) A^T\big),
\end{equation}
with the employed seminorm induced by the positive semidefinite matrix $A^T \!A$ for a given $A \in \R^{l \times n}$.

To solve the minimization problem \eqref{eq:Aoptimal} and to find the optimal design for the $(k+1)$th X-ray projection, we resort to the exhaustive optimization algorithm introduced in \cite{Burger21}. In our numerical experiments, the weight $A$ is always the identity matrix ${\rm I} \in \R^{n \times n}$, that is, we consider the reconstruction accuracy equally important at all pixels/voxels. If one were only interested in the accuracy of the reconstruction inside a certain
region of interest, one could select $A = {\rm I}_{\rm ROI} \in  \R^{n \times n}$ having ones at the diagonal positions corresponding to the pixels/voxels in the
region of interest and zeros as its all other elements~\cite{Burger21}.

\begin{remark}
  \label{rm:discr}
  Finding the optimal design parameter via \eqref{eq:Aoptimal} is computationally more demanding than computing an edge-enhancing reconstruction using the lagged diffusivity ideas presented in Section~\ref{sec:lagged_diff}. However, one can speed up the optimization step by implementing it using a sparser discretization than the one employed for computing the actual reconstructions: Once the reconstruction $u_{k} \in \R^n$ corresponding to the first $k$ projection images has become available, it is interpolated onto a sparser grid with $\tilde{n} \leq n$ interior pixels/voxels to obtain $\tilde{u}_k \in \R^{\tilde{n}}$. The corresponding covariance matrix $\tilde{\Gamma}_k$ is then formed as in \eqref{eq:posterior_cov} but with $\mathbf{R}_k$ replaced by the analogous (stacked) X-ray projection matrix for the sparser discretization and with $\Gamma_{k-1}^{(J)}$ replaced by $H(\tilde{u}_{k})^{-1}$ formed as in \eqref{eq:H} but using a Lagrangian finite element basis for the sparser discretization. The (approximate) posterior for the interpolated absorption $\tilde{\Gamma}_{\rm post}^{(k+1)}(p)$ is then as in \eqref{eq:p_postcov} but with $\Gamma_k$ replaced by $\tilde{\Gamma}_k$ and $R(p)$ with an X-ray projection matrix corresponding to the sparser discretization. Finally, the optimal design parameters (that are discretization invariant in our numerical experiments) are computed via \eqref{eq:Aoptimal} with $\Gamma_{\rm post}^{(k+1)}(p) \in \R^{n \times n}$ replaced by $\tilde{\Gamma}_{\rm post}^{(k+1)}(p) \in \R^{\tilde{n} \times \tilde{n}}$ and the weight matrix $A$ modified appropriately.
  \end{remark}

\section{Sequential edge-promoting optimization of projections}
\label{sec:opti}

In this section, the above developments are summarized by combining the lagged diffusivity iteration and the sequential optimization of X-ray projections into a single concise algorithm. In the following it is assumed that the overall number of X-ray projections $K \in \N$ is known in advance, but in practice the operator of the algorithm can stop the iteration as soon as the reconstruction is considered good enough, thus treating $K$ as the {\em maximum} number of projection images.

\begin{algorithm}
\label{alg:basic_optimization}
\begin{algorithmic}
  \STATE{Select the prior parameters $T>0$ and $\gamma >0$, a tolerance for the stopping criterion $\tau > 0$, the number of iterations $K \in \N$, and the weight matrix $A$.
    }

\vspace{1.5mm}

  \STATE{Initialization:}
  \STATE{\hspace{2mm} $\rhd$ Set $u_0 = \mathbf{1} \in \R^n$.}
  \STATE{\hspace{2mm} $\rhd$ Define $\Gamma_{0} := H(u_0)^{-1}$ according to \eqref{eq:H}.}

\vspace{1.5mm}

  \STATE{Iteration:}

  \FOR{$k=1, \dots, K$}
  \STATE{\hspace{2mm} $\rhd$ Solve for $p_{k}$ via \eqref{eq:Aoptimal} with $\Gamma_{\rm post}^{(k)}(p)$ defined by \eqref{eq:p_postcov}~\cite{Burger21}.}
  \STATE{\hspace{2mm} $\rhd$ Form the projection matrix $R(p_k)$ and `measure' the data $y_k$.}
  \STATE{\hspace{2mm} $\rhd$ Set $j=0$, $u_{k-1}^{(0)}=u_{k-1}$, and $\Delta \Phi = \tau + 1$.}
  \WHILE{$\Delta \Phi > \tau$}
  \STATE{\hspace{2mm} $\rhd$ Set $j \leftarrow j+1$.}
     \STATE{\hspace{2mm} $\rhd$ Form $\Gamma_{k-1}^{(j)}$ according to \eqref{eq:step1}.}
     \STATE{\hspace{2mm} $\rhd$ Compute $u_{k-1}^{(j)}$ according to \eqref{eq:step2}.}
     \STATE{\hspace{2mm} $\rhd$ Compute $\Delta \Phi = |\Phi(u_{k-1}^{(j-1)}) - \Phi(u_{k-1}^{(j)})|/\Phi(u_{k-1}^{(j)})$.}
  \ENDWHILE
  \STATE{\hspace{2mm} $\rhd$ Define $\Gamma_{k} = \Gamma_{k-1}^{(j)}$ and $u_k = u_{k-1}^{(j)}$.}
  \ENDFOR

  \vspace{2mm}

  \RETURN $u_{K}$ and $\Gamma_{K}$.
\end{algorithmic}
\end{algorithm}

The stopping criterion for the interior loop is motivated by material in \cite{Arridge13}: Apart from the case $j=0$, the value of the (smoothened) TV functional $\Phi$ typically decreases monotonically during the lagged diffusivity iteration because the reconstruction becomes gradually better aligned with the prior information. The iteration is stopped once the relative convergence rate falls below a preselected tolerance $\tau > 0$.

In many of the following numerical examples, the deduction of the sequentially A-optimal projections,~i.e.~the first step in the exterior loop of Algorithm~\ref{alg:basic_optimization}, is performed on a sparser discretization of $D$ consisting of $\tilde{n} < n$ interior pixels/voxels in order to speed up the computations. The modifications required by this accelerated algorithm are described in Remark~\ref{rm:discr}. Consult \cite{Burger21} for more information on the exhaustive algorithm for defining the optimal projections.

\section{Numerical experiments}
\label{sec:numerics}
Both two and three-dimensional numerical examples are presented. In all tests, the free parameters in Algorithm~\ref{alg:basic_optimization} are chosen as $T = 10^{-6}$, $\gamma = 10^{-2}$, $\tau = 10^{-4}$ and $A = \mathrm{I}$. The algorithm is not very sensitive to the chosen (reasonably small) value for $T$. The other two parameters $\gamma$ and $\tau$ do affect the numerical results, but as our main aim is to compare reconstructions with and without sequential optimization of the projection geometries, we do not dwell on their selection. The choice of $A$ reflects that we are equally interested in the reconstruction quality everywhere in $D$. The components of the additive zero-mean Gaussian noise contaminating the measurements are assumed to be mutually independent with a common standard deviation $\sigma > 0$ that may vary between the experiments. In other words, all noise covariance matrices appearing in
Sections~\ref{sec:lagged_diff} and \ref{sec:A_optimality} are assumed to be of the form $\sigma^2 \mathrm{I}$, with $\mathrm{I}$ being an identity matrix of the appropriate size.

When the performance of Algorithm~\ref{alg:basic_optimization} is compared to reconstructions corresponding to, say, equiangular projections, the reference reconstructions are computed via a single lagged diffusivity iteration with the same, aforelisted values for the parameters $T$, $\gamma$, $\tau$ and $A$. To be more precise, if $\mathbf{R}$ is the projection matrix corresponding to all employed reference geometries, $\mathbf{y}$ is the corresponding data vector and $\bm{\Gamma}_{\rm noise} = \sigma^2 \mathrm{I}$ is the assumed noise covariance, then one starts from the initial guess $u^{(0)} = \mathbf{1} \in \R^n$ and iterates the two steps
 $$
 \Gamma^{(j)} = H(u^{(j-1)})^{-1}, \qquad
 u^{(j)} =  \Gamma^{(j)} \mathbf{R}^T \big(\mathbf{R} \Gamma^{(j)} \mathbf{R}^T + \gamma \bm{\Gamma}_{\rm noise}  \big)^{-1}\mathbf{y}
 $$
 until
 $$
\frac{\big|\Phi(u^{(j-1)}) - \Phi(u^{(j)})\big|}{\Phi(u^{(j)})} < \tau,
$$
after which $u^{(j)}$ is dubbed the reconstruction.
In other words, one essentially runs the interior loop of Algorithm~\ref{alg:basic_optimization} assuming that all (equiangular) projection geometries and the associated data are available to start with.

\begin{remark}
We do not claim that the lagged diffusivity iteration is the best method for computing TV type reconstructions in X-ray tomography. However, since the algorithm for deducing the optimal projection geometries is inherently connected to the lagged diffusivity ideas, we consider using a simplified version of Algorithm~\ref{alg:basic_optimization} for computing the control reconstructions corresponding to nonoptimized projection geometries a well motivated choice. In particular, this enables focusing solely on the effect of the optimal design when comparing the reconstructions.
  \end{remark}

\begin{figure}[t]
	\centering
\centering
\begin{tikzpicture}[scale=1.2]
  \draw[->, thick] (-1.2, -1.2) -- (2.5, -1.2);
  \draw[->, thick] (-1.2, -1.2) -- (-1.2, 2.5);

  \draw[dashed] (0,0) circle[radius=2.12];

  \draw[very thin, color=gray, step=0.2cm] (-1.2,-1.2) grid (1.2,1.2);

  \draw (0.55,-2.45) -- (2.45,-0.55);
  \draw [->] (-2.15,0.85) node[anchor=south east] {Source} -- (0.85,-2.15);
  \draw [->] (-1.7167,1.2833) -- (1.2833,-1.7167);
  \draw [->] (-1.2833,1.7167) -- (1.7167,-1.2833) node[anchor=north west] {Detectors};
  \draw [->] (-0.85,2.15) -- (2.15,-0.85);

  \draw[line width=0.5mm] (0.75,-2.25) -- (0.95,-2.05);
  \draw[line width=0.5mm] (1.1833,-1.8167) -- (1.3833,-1.6167);
  \draw[line width=0.5mm] (1.6167,-1.3833) -- (1.8167,-1.1833);
  \draw[line width=0.5mm] (2.05,-0.95) -- (2.25,-0.75);
  \draw[line width=0.2cm, color=red] (-2.25,0.75) -- (-0.75,2.25);
  \node[] at (1.0, 1.4) {Object};
\end{tikzpicture}
	\caption{Two-dimensional measurement setup.}
        \label{fig:2Dgeometry}
\end{figure}
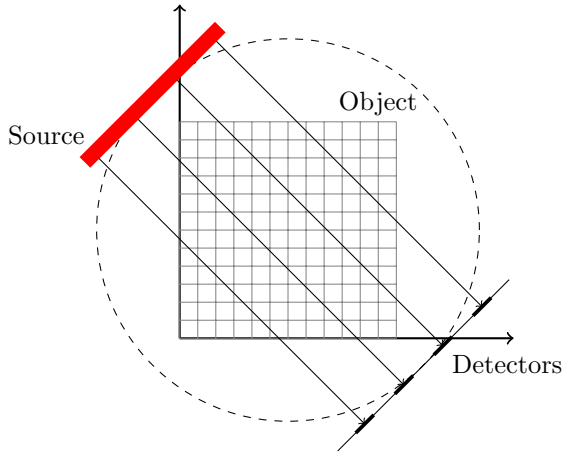

\subsection{Two-dimensional parallel beam tomography}
In our two-dimensional numerical experiments, the measurement setup is the same as described in \cite{Burger21}. That is, the domain $D = [0,1]^2$ is discretized into $n = N^2$ square pixels, through which we take projections consisting of a number of parallel X-rays; see Figure~\ref{fig:2Dgeometry}. The individual X-rays are equally spaced and have a fixed width for a particular experiment. The width of the whole source-receiver pair satisfies $0< w \leq 1$. The components of the design variable $p \in \R^2$ for a single projection geometry define the projection angle and the signed distance from the center of $D$ to the median line of the source-receiver pair. The latter component of $p$ is restricted within the interval $[w-1, 1-w]/2$.

We present three two-dimensional experiments. The first one exemplifies the general behavior of Algorithm~\ref{alg:basic_optimization} with a simple target. The effect of optimizing the projection geometries on a sparser grid than the one used for forming the reconstructions is also tested; see Remark~\ref{rm:discr}. In the second test, the superiority of Algorithm~\ref{alg:basic_optimization} over the usage of equiangular full-width projections with an equivalent radiation dose is statistically demonstrated in the case of certain randomly selected phantoms. Finally, the third test applies Algorithm~\ref{alg:basic_optimization} to the Shepp--Logan phantom.

\subsubsection{2D~Test~1: Explicit example with a simple target}
\begin{figure}
	\centering
  \includegraphics[width = 0.49\columnwidth]{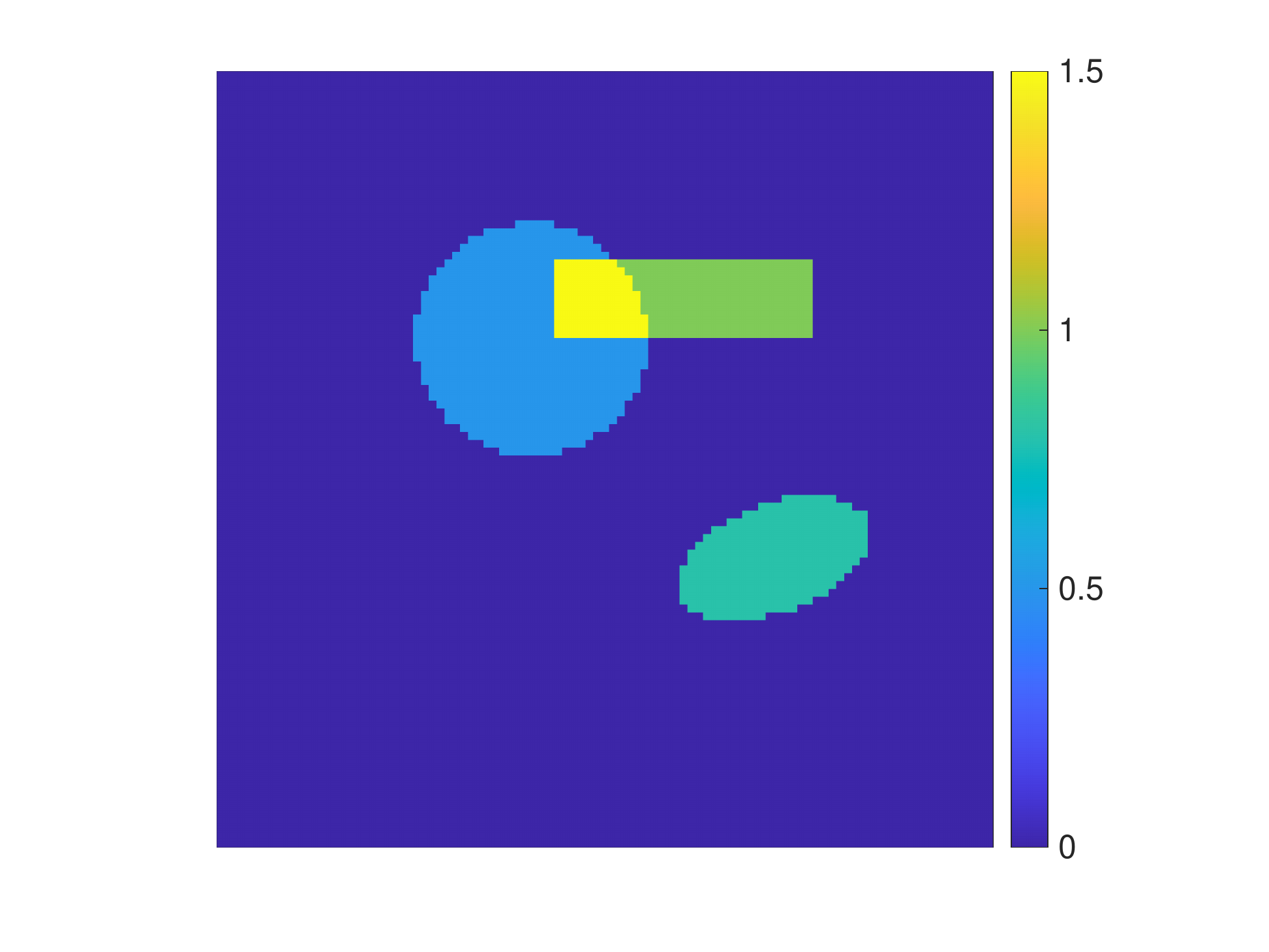}
  \includegraphics[width = 0.49\columnwidth]{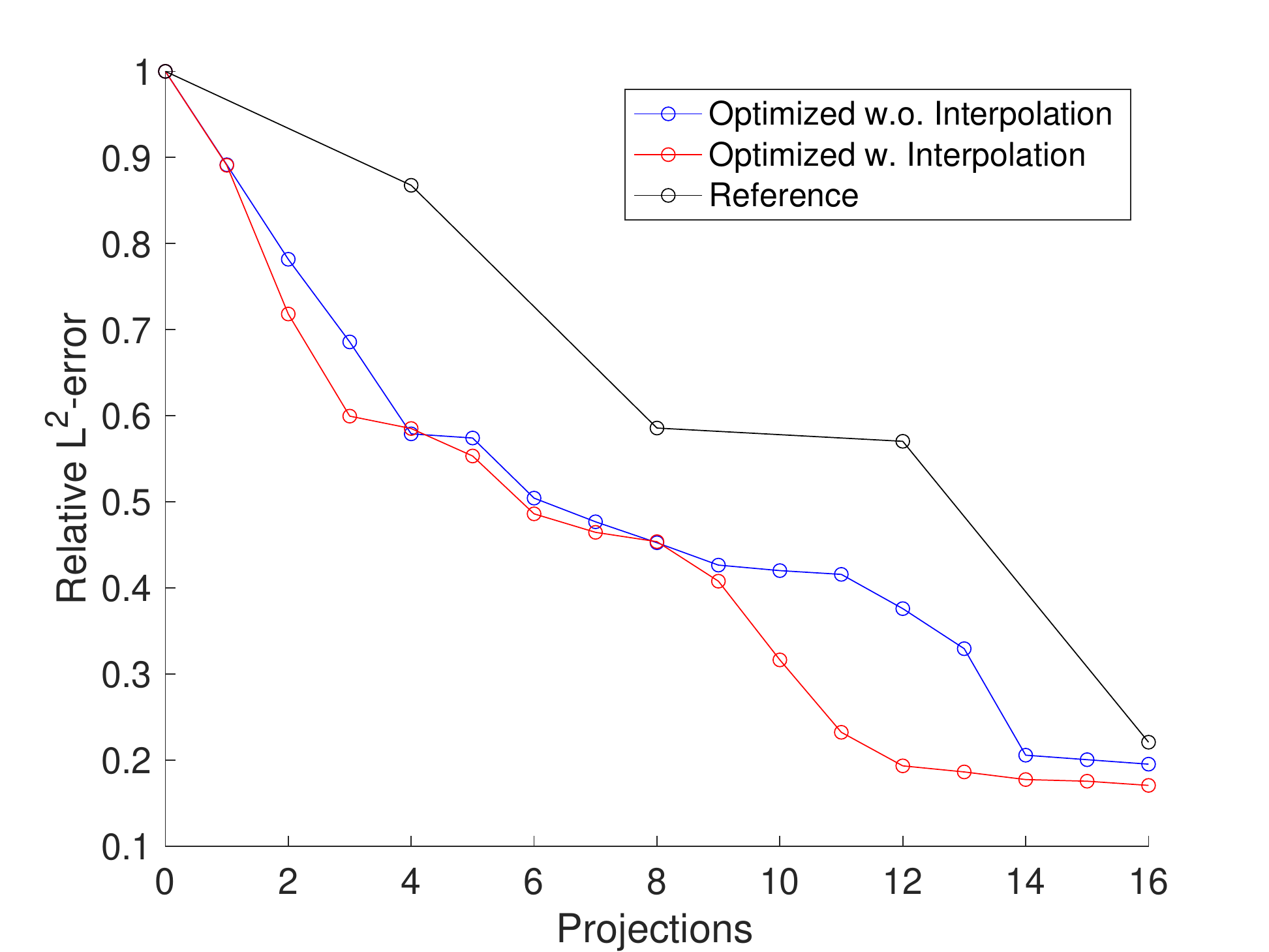}
	\caption{{\sc 2D~Test~1.} Left: Target with the rectangle, circle and ellipse having absorption levels $1$, $0.5$, and $0.8$ respectively. Right: Relative $L^2(D)$ errors for the reconstructions. The blue curve corresponds to projections optimized with the dense discretization for $D$, the red curve to projections optimized with the sparse discretization for $D$, and the black curve to the equiangular full-width reference projections. The horizontal axis indicates the number of projections with the beam width $0.25$.}
	\label{fig:test1phantom}
\end{figure}

The aim of our first numerical experiment is to demonstrate the basic functioning of Algorithm~\ref{alg:basic_optimization}. The target, shown in the left-hand image of Figure \ref{fig:test1phantom}, consists of three simple shapes, each with a different uniform absorption level, placed randomly inside $D$. The absorption of the background is zero. The target has $N = 100$ pixels per edge, and the number of individual sensors in a full-width source-receiver pair is $51$. The noise level is set to $\sigma = 10^{-3}$, which corresponds to a noise-to-signal ratio of at least $0.2$\% for all line integrals considered in the inversion. The beam width is chosen to be $0.25$, which is a quarter of the maximal source-receiver pair width and corresponds to $13$ individual X-rays. Algorithm~\ref{alg:basic_optimization} is run for a total of $K=16$ iterations. In addition to considering the basic form of Algorithm~\ref{alg:basic_optimization}, we also test speeding up the computations by performing the selection of the projection geometries on a considerably sparser discretization of the domain with only $\tilde{N} = 31$ pixels per edge; see Remark~\ref{rm:discr} for more details and note, in particular, that the actual reconstructions are still formed on the denser grid with $N^2$ pixels. For comparison, we also compute reconstructions from equiangular full-width projections corresponding to equivalent radiation doses.

The right-hand image of Figure~\ref{fig:test1phantom} shows the relative $L^2(D)$ reconstruction errors after each step of Algorithm~\ref{alg:basic_optimization};
the blue curve corresponds to optimizing the projection geometries on the reconstruction grid with $10^4$ pixels and the red curve to performing the optimization steps of the algorithm using the considerably sparser discretization with only $\tilde{N}^2 \approx 10^3$ pixels. The black line depicts the relative $L^2(D)$ errors resulting from the equiangular reference projections. Note that one projection with the maximal beam width of $1$
approximately corresponds to the same amount of data, or equivalently the same radiation dose, as four projections with the beam width $0.25$. As a consequence, the labels at $4$, $8$, $12$ and $16$ on the horizontal axis correspond to one, two, three and four equiangular reference projections, respectively.

\begin{figure}
	\centering
  \includegraphics[width = 0.31\columnwidth]{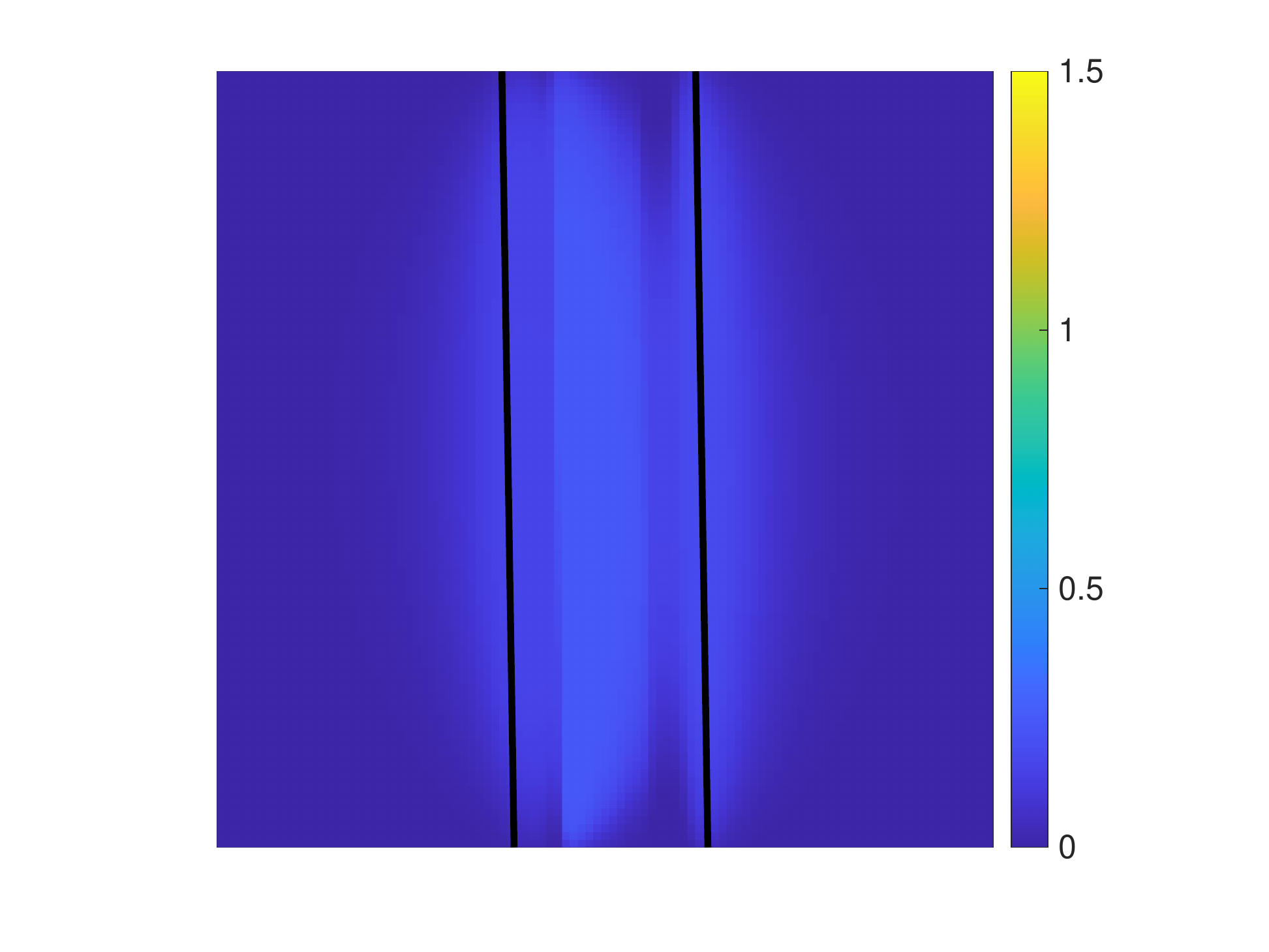}
  \includegraphics[width = 0.31\columnwidth]{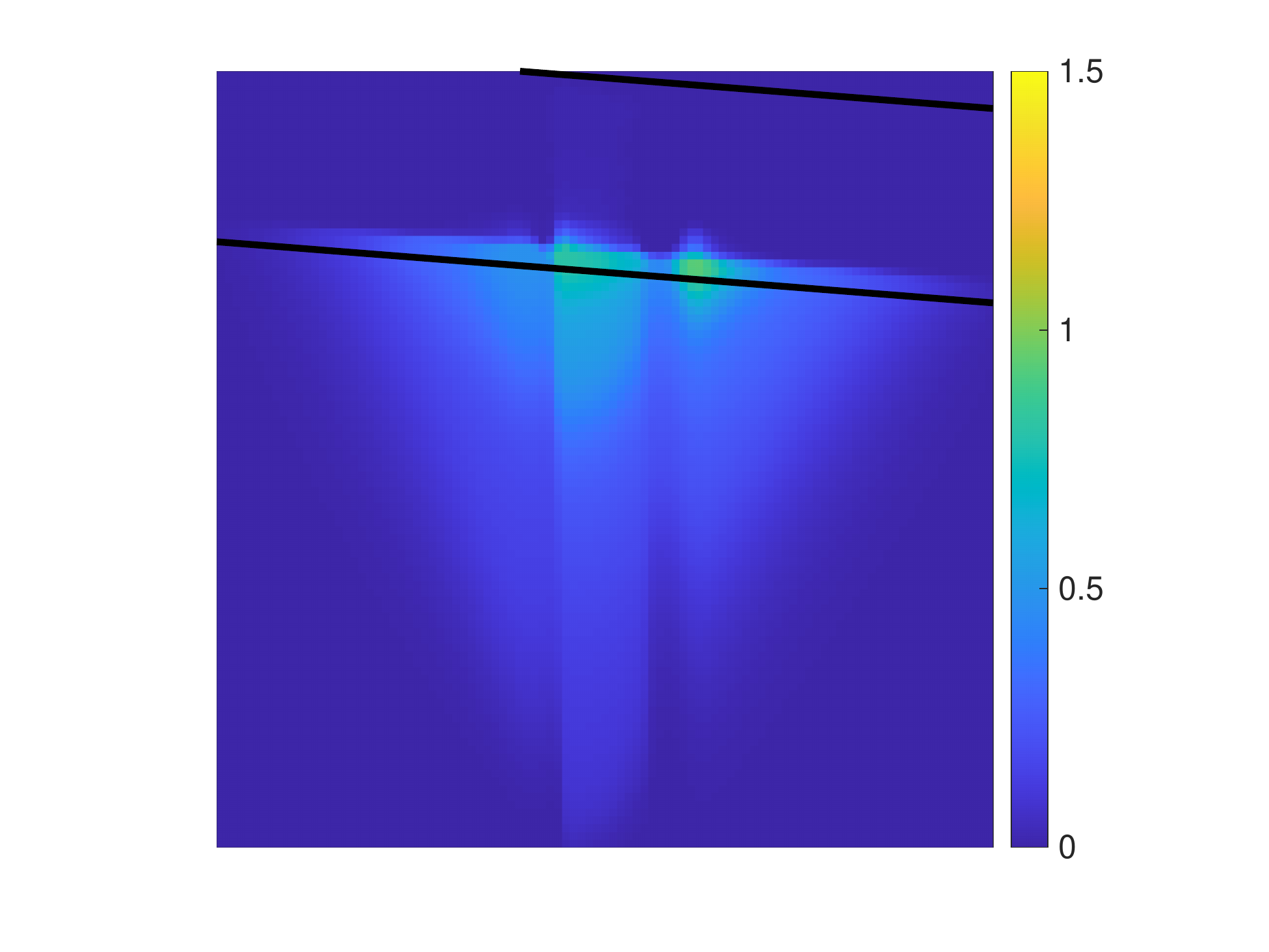}
  \includegraphics[width = 0.31\columnwidth]{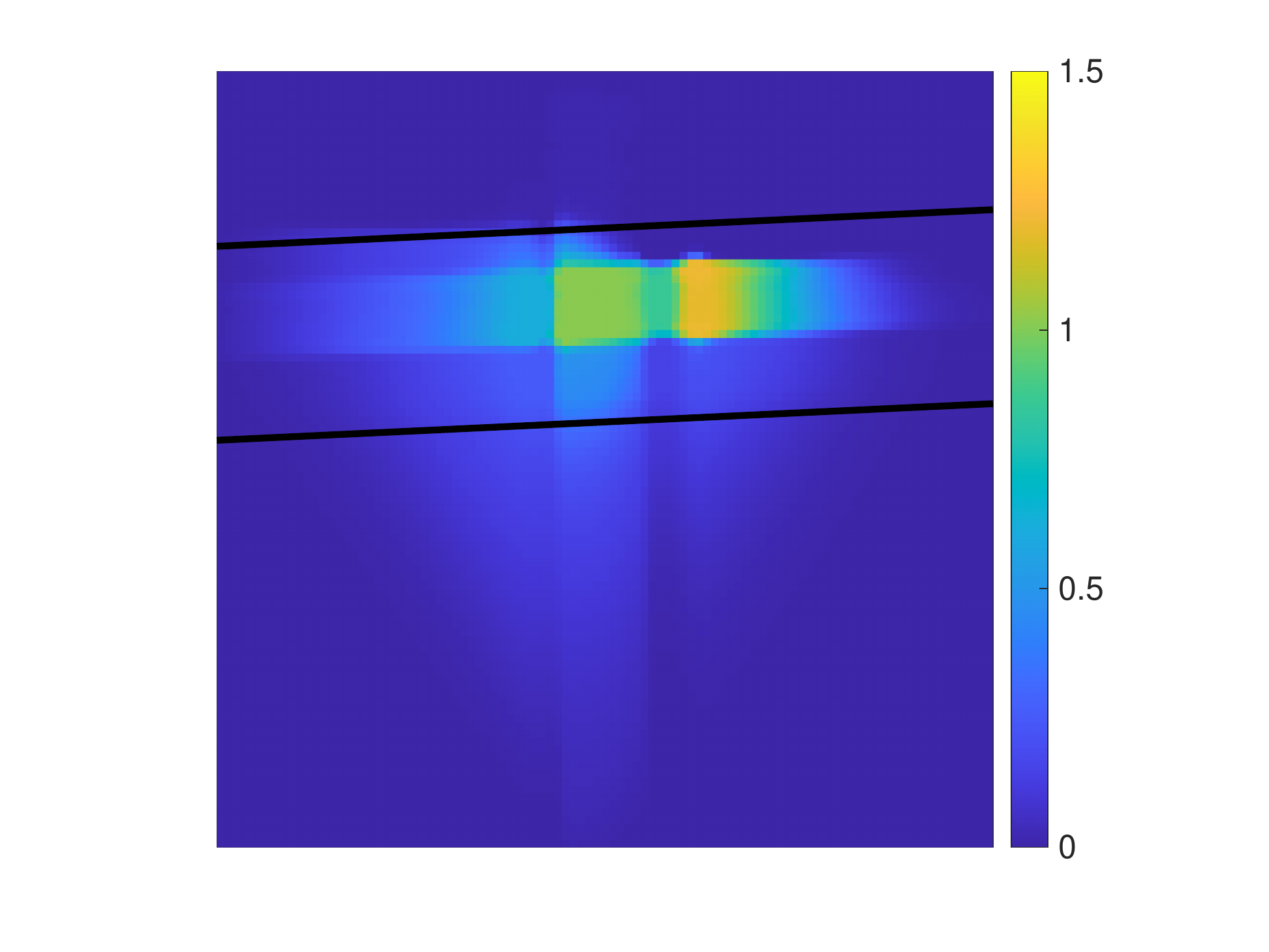}
  \includegraphics[width = 0.31\columnwidth]{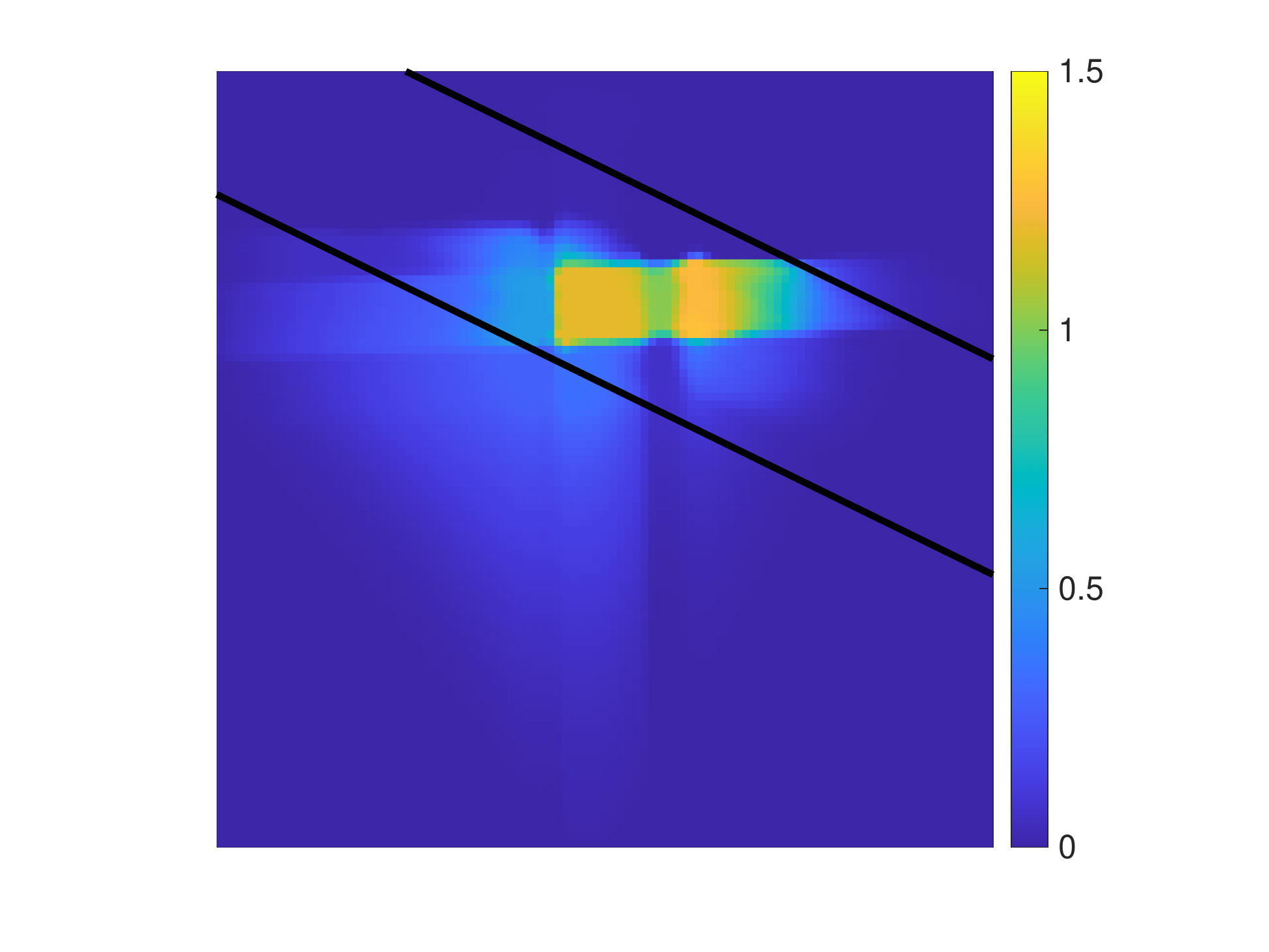}
  \includegraphics[width = 0.31\columnwidth]{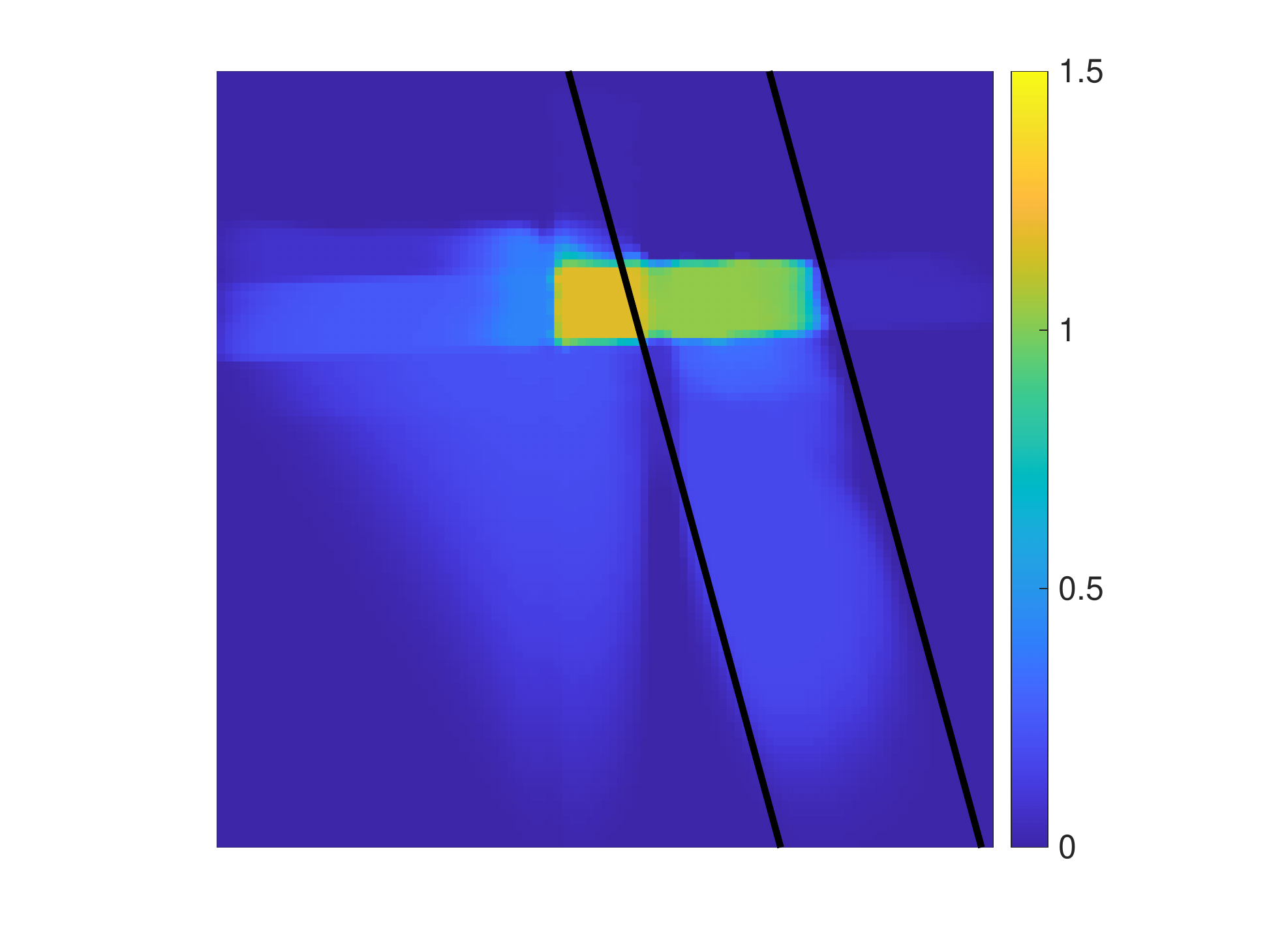}
  \includegraphics[width = 0.31\columnwidth]{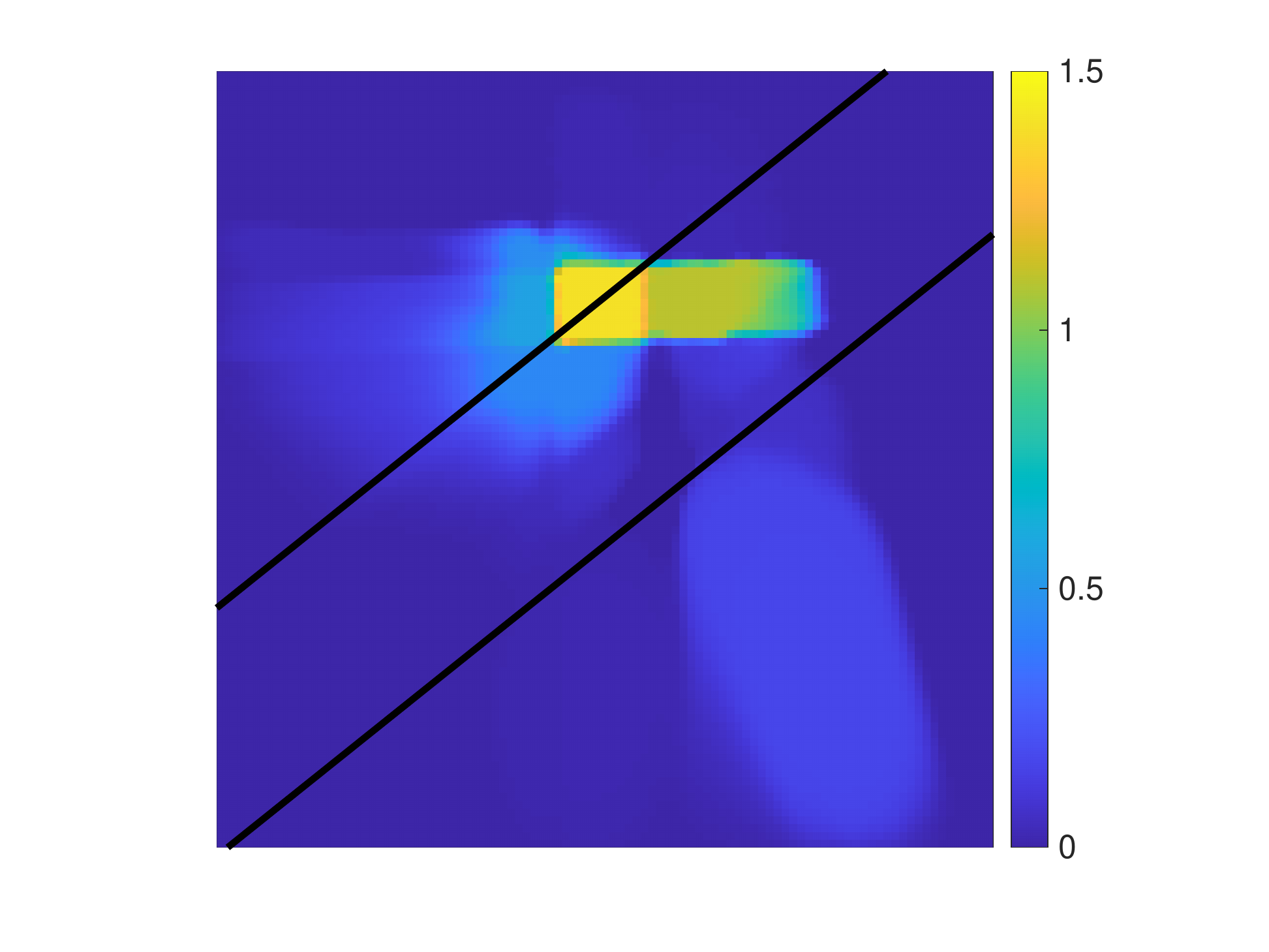}
  \includegraphics[width = 0.31\columnwidth]{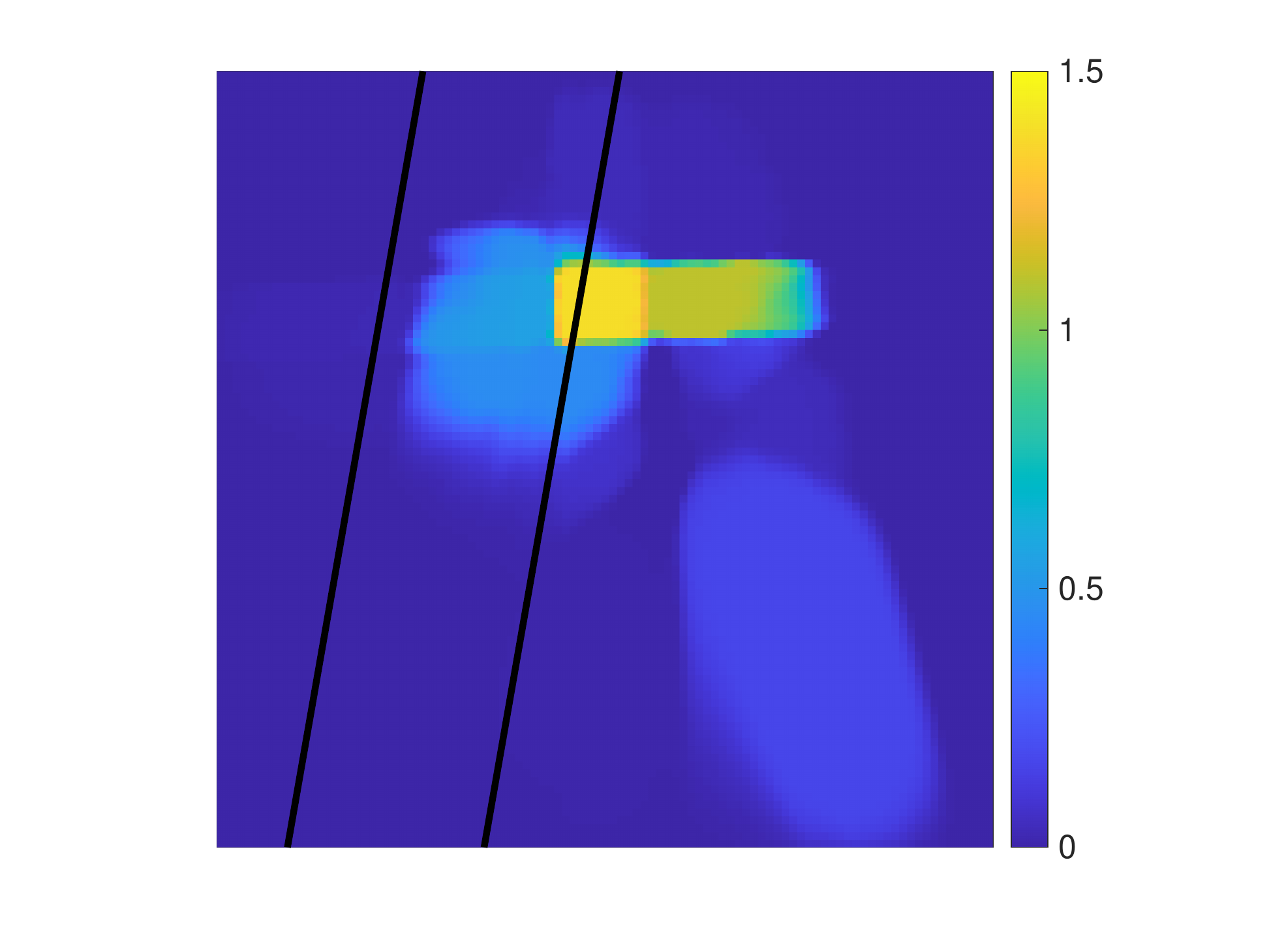}
  \includegraphics[width = 0.31\columnwidth]{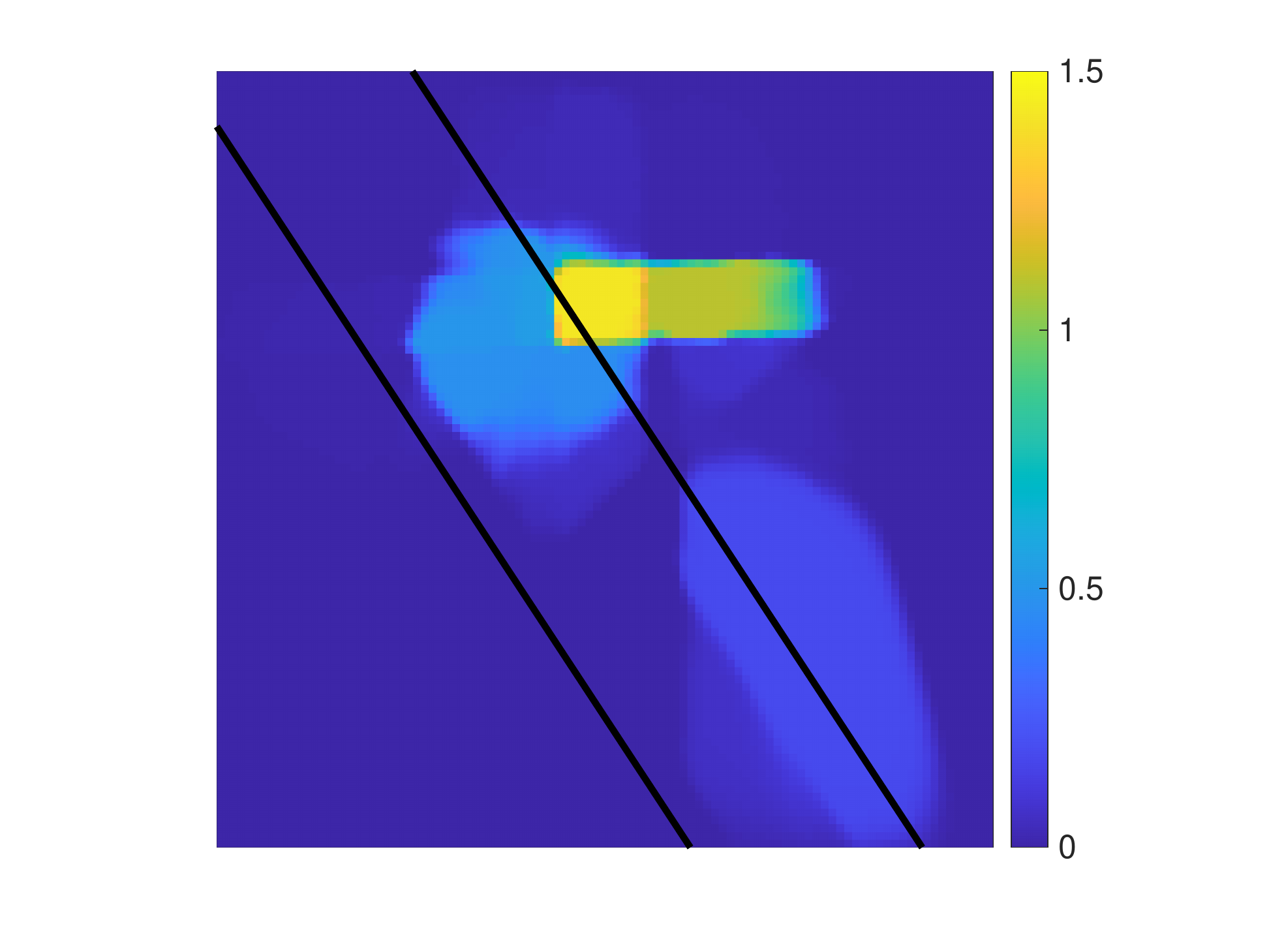}
  \includegraphics[width = 0.31\columnwidth]{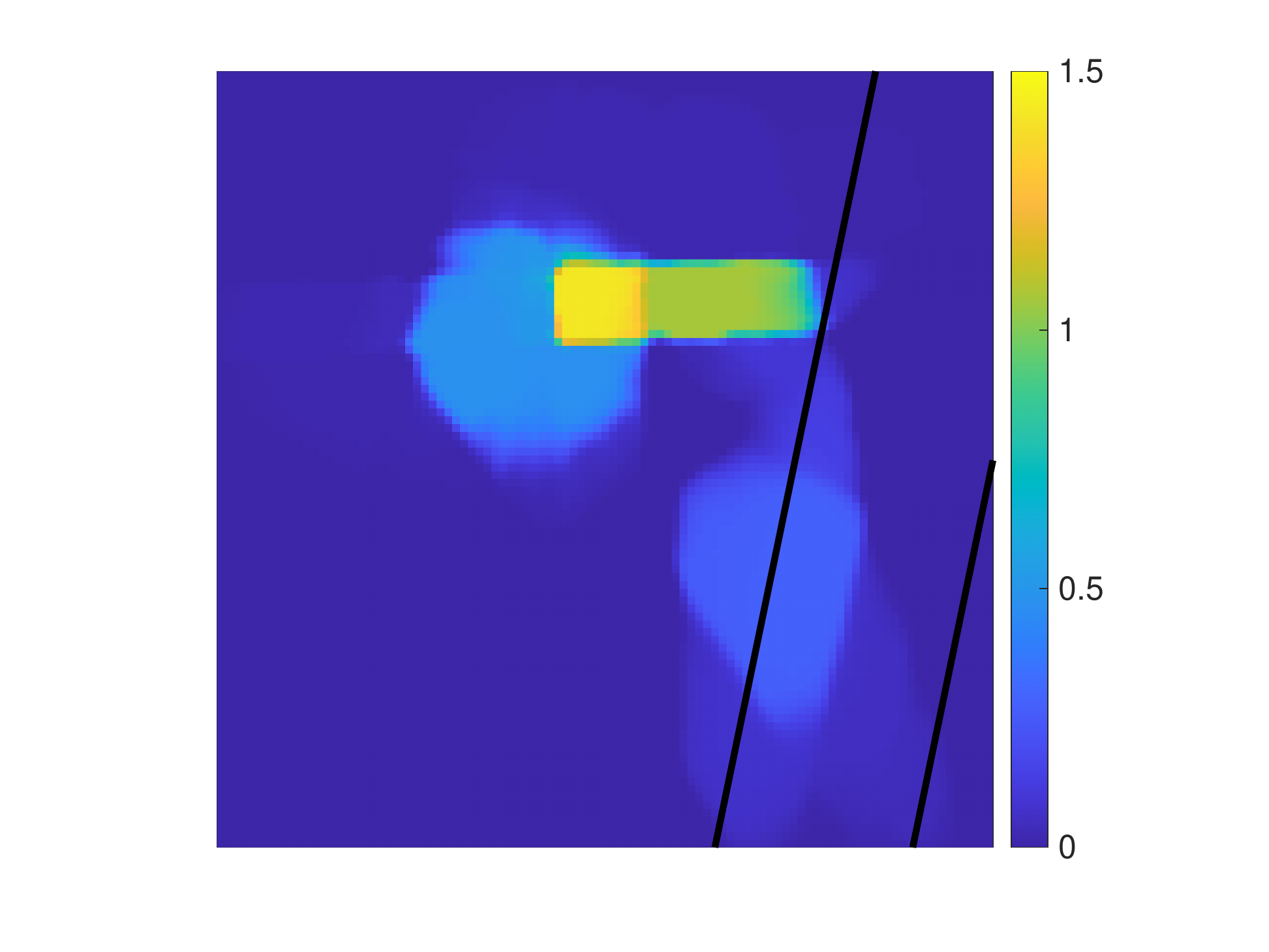}
  \includegraphics[width = 0.31\columnwidth]{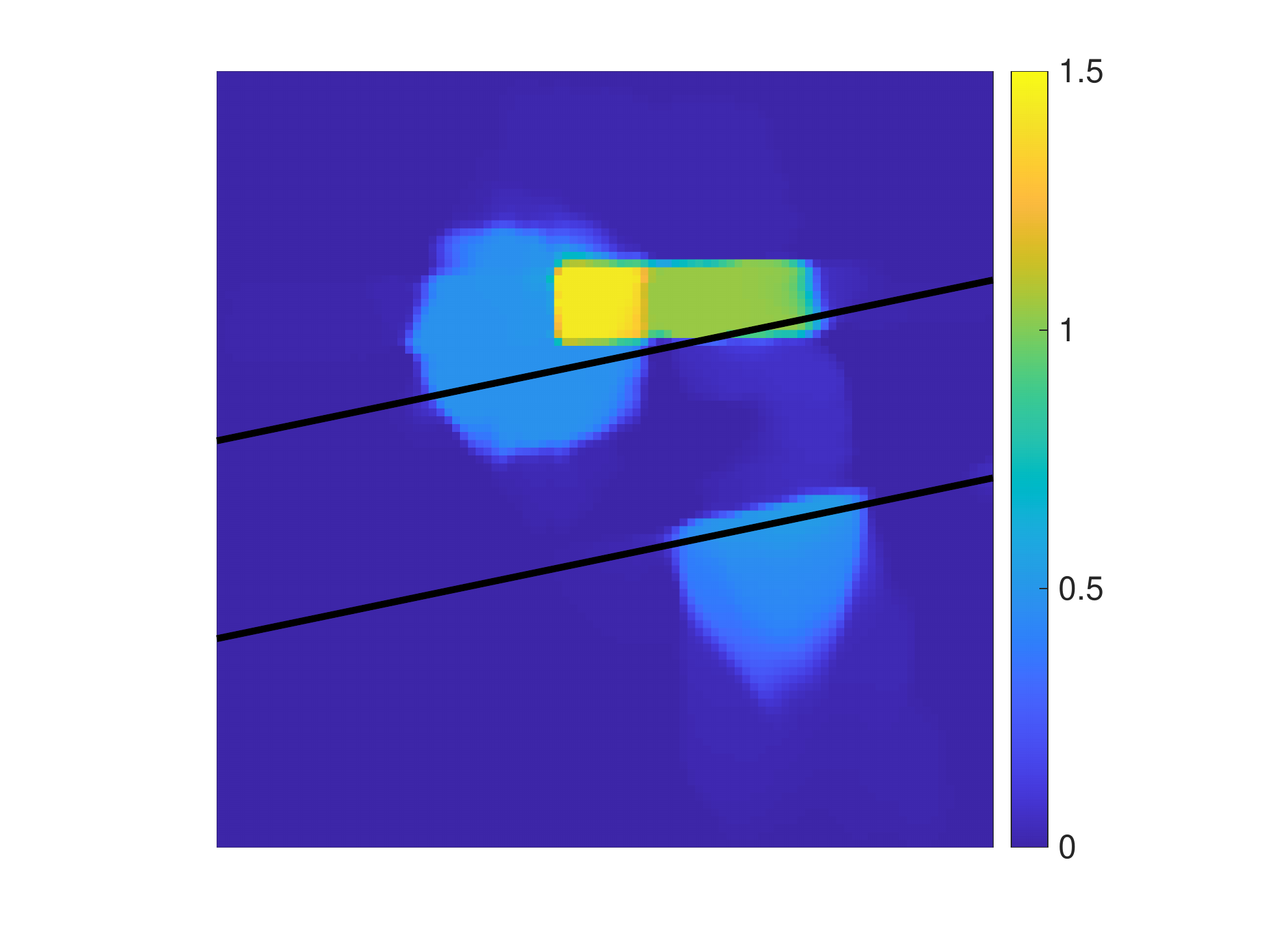}
  \includegraphics[width = 0.31\columnwidth]{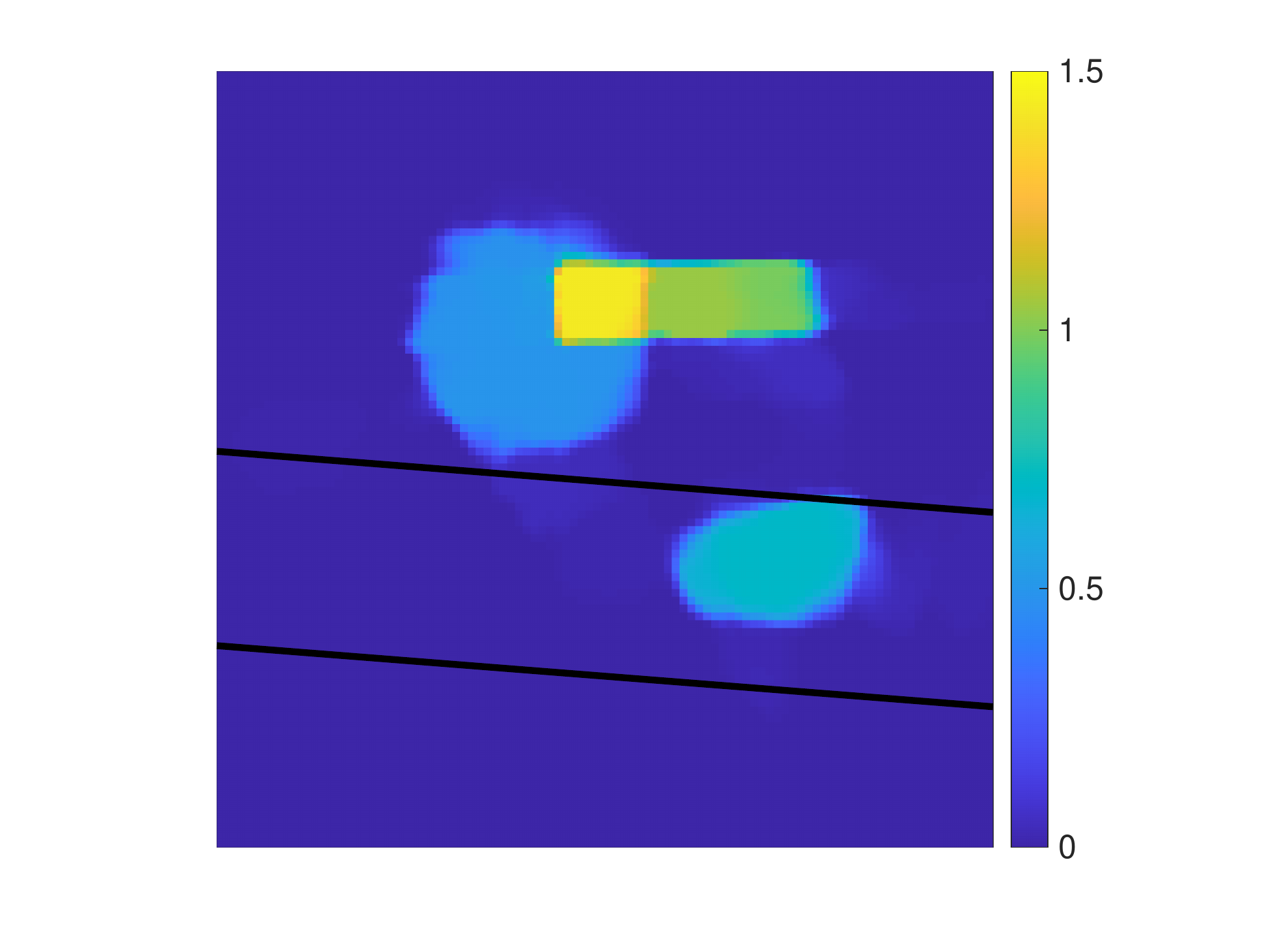}
   \includegraphics[width = 0.31\columnwidth]{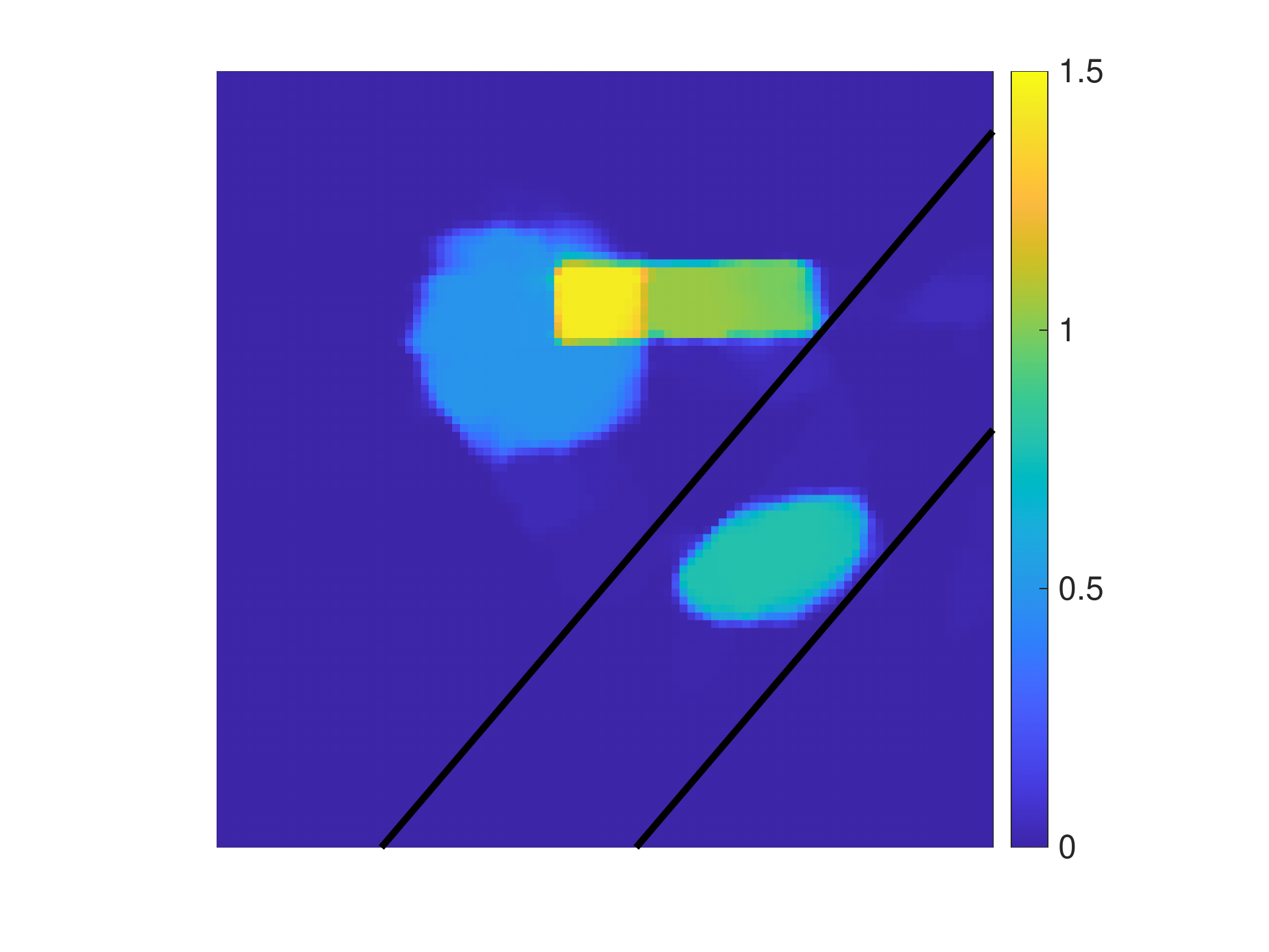}
	\caption{{\sc 2D~Test~1.} First twelve optimized projection geometries and the corresponding reconstructions.}
	\label{fig:test1reconstructions}
\end{figure}

According to Figure~\ref{fig:test1phantom}, the $L^2(D)$ reconstruction errors at equivalent radiation doses are lower for the sequentially optimized projection geometries with the quarter-width source-receiver pair than for the equiangular full-width projections. This is not very surprising as the full-width projections (are forced to) waste radiation to image regions that contain nothing interesting, whereas the optimized projections concentrate on areas of interest; cf.~Figure~\ref{fig:test1reconstructions}. On the other hand, deducing the optimal designs employing the sparser discretization for $D$ does not seem to considerably hamper the overall performance of Algorithm~\ref{alg:basic_optimization}, although the discretization level does affect the precise specifications of the individual optimized projection geometries. After sufficiently many projections, the advantage of Algorithm~\ref{alg:basic_optimization} over the equiangular full-width projections becomes almost negligible.

\begin{figure}
	\centering
  \includegraphics[width = 0.31\columnwidth]{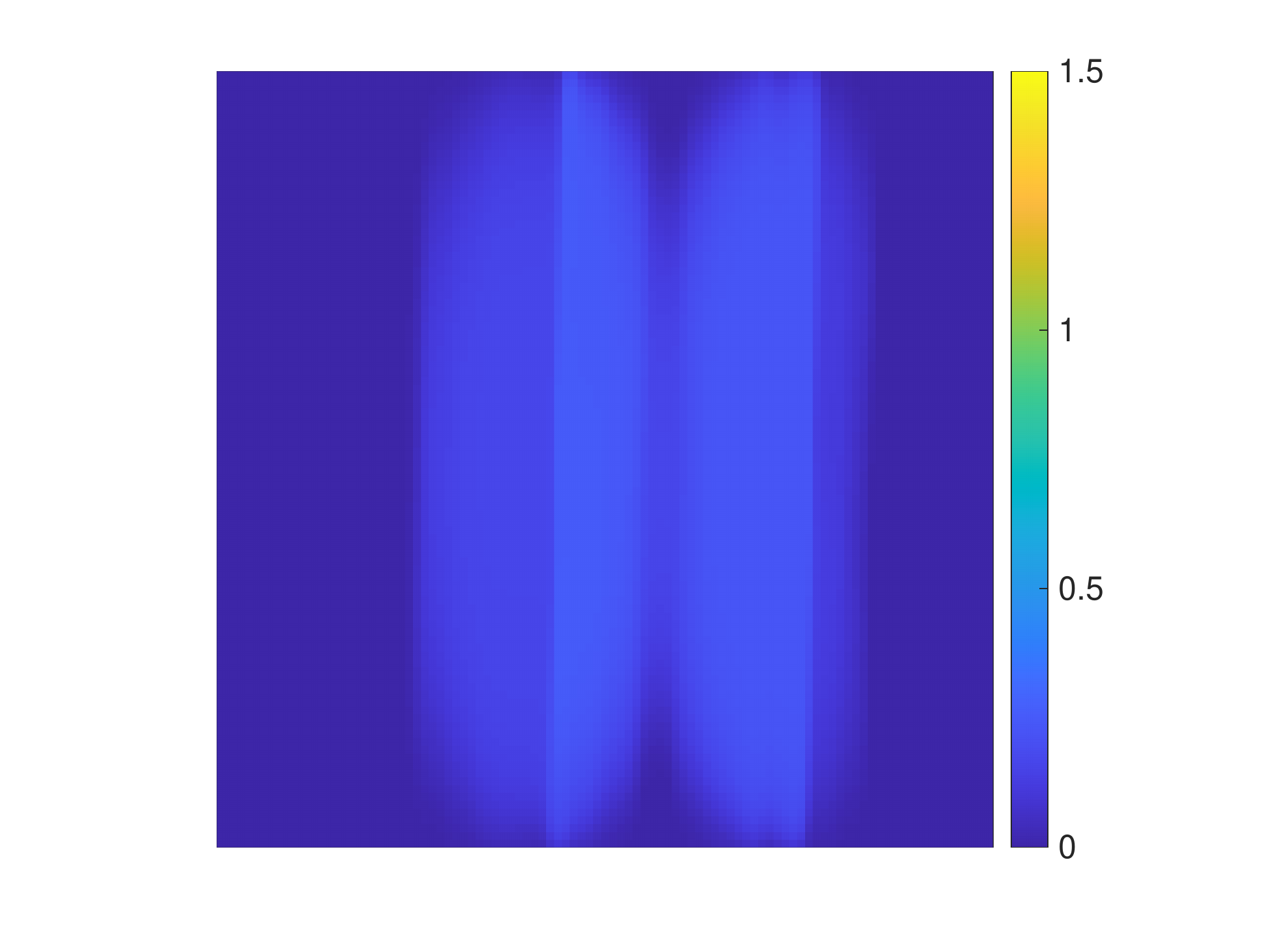}
  \includegraphics[width = 0.31\columnwidth]{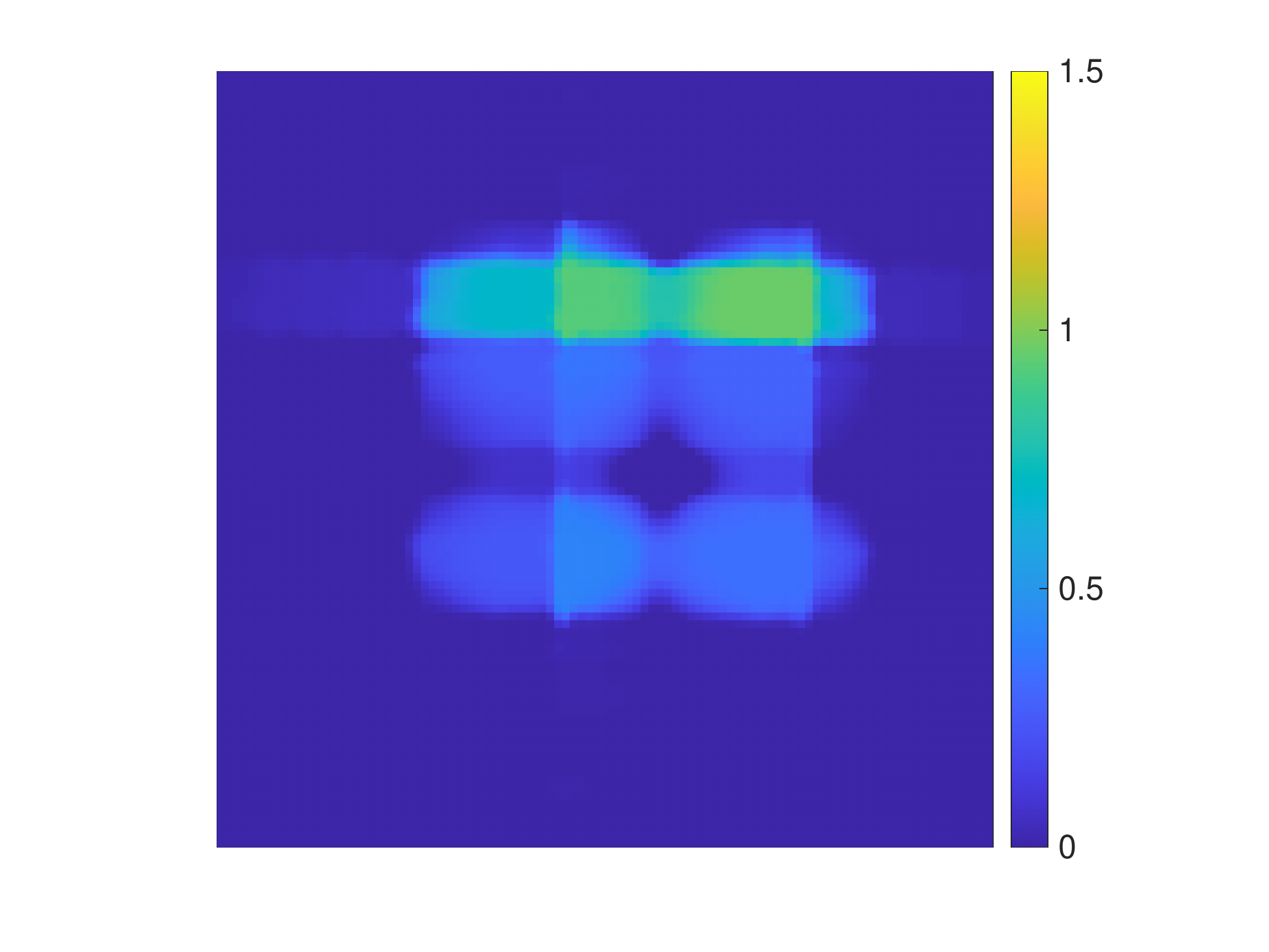}
  \includegraphics[width = 0.31\columnwidth]{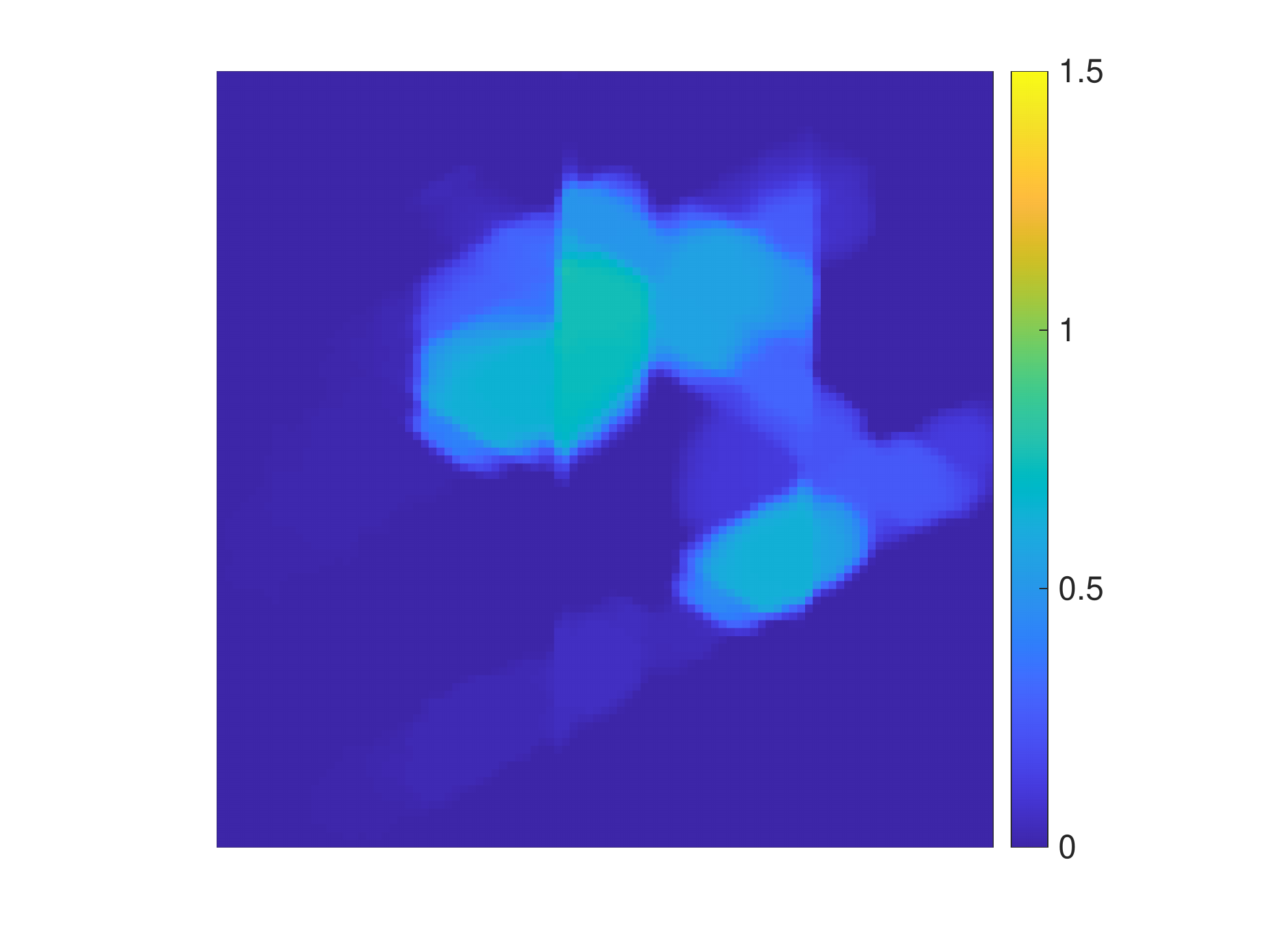}
	\caption{{\sc 2D~Test~1.} Reference reconstructions for one, two and three equiangular full-width projections.}
	\label{fig:test1reconstructions_ref}
\end{figure}

Figure~\ref{fig:test1reconstructions} shows the projection geometries and the corresponding reconstructions for the first $12$ iterations of Algorithm~\ref{alg:basic_optimization}. At least in the considered setup, the algorithm does indeed seem to have a tendency to concentrate the projections over areas where the reconstruction already shows quick variations. Occasionally other areas are also explored, cf.~the ninth projection. After ten iterations all target shapes are already clearly visible. For comparison, Figure~\ref{fig:test1reconstructions_ref} shows the first three reconstructions corresponding to the full-width equiangular reference projections.

\subsubsection{2D~Test~2: Average errors over random targets}
In the second numerical experiment, the aim is to statistically demonstrate that Algorithm~\ref{alg:basic_optimization} has the potential to produce on average better reconstructions for a limited radiation dose than a straightforward approach with equiangular full-width projections. To this end, the algorithm is run with the beam widths of $0.25$ and $0.5$ for a set of random targets, and the average relative $L^2(D)$ reconstruction errors are compared to those obtained by the equiangular approach.

\begin{figure}
	\centering
	\includegraphics[width = 0.45\columnwidth]{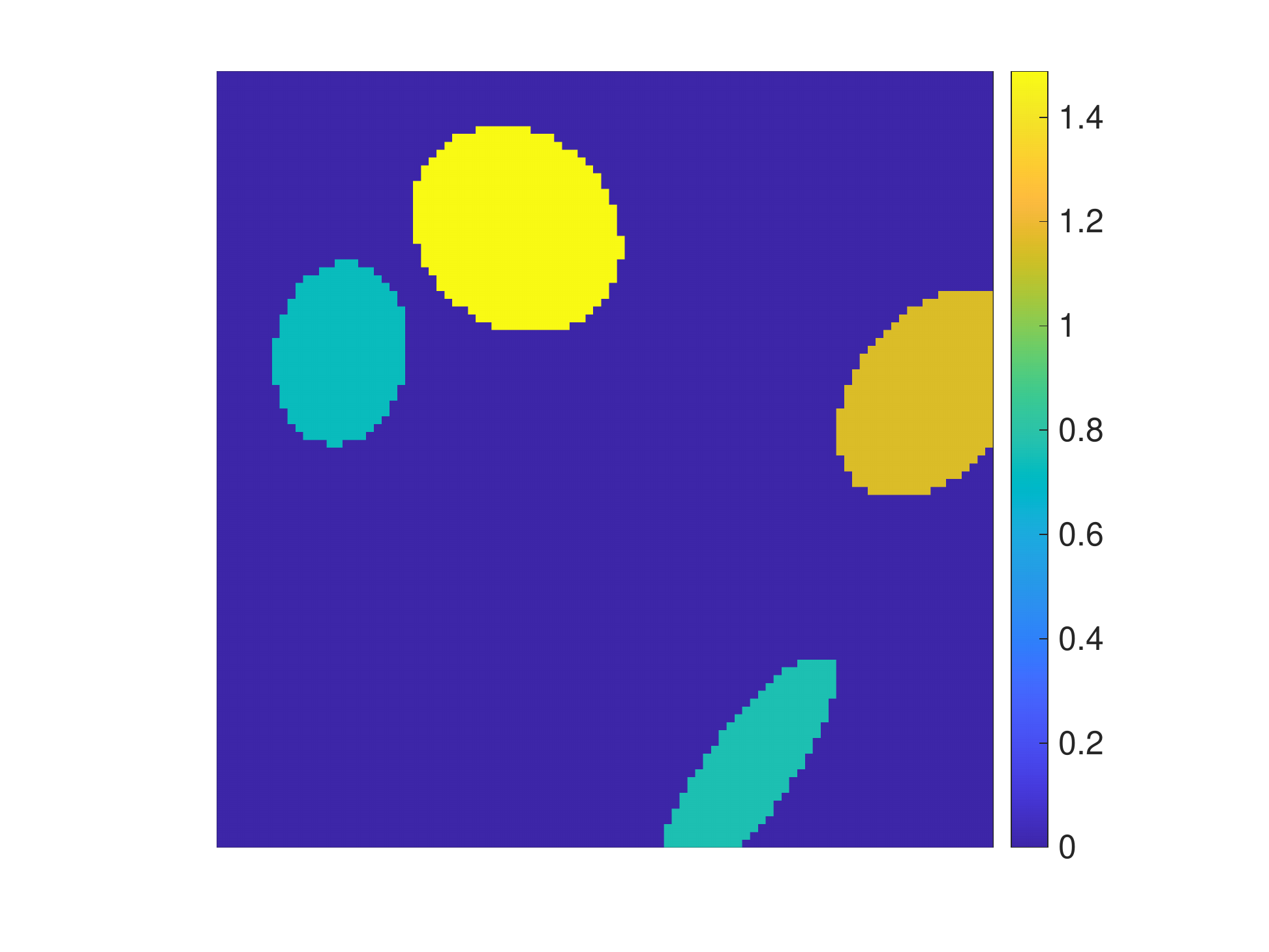}
	\includegraphics[width = 0.45\columnwidth]{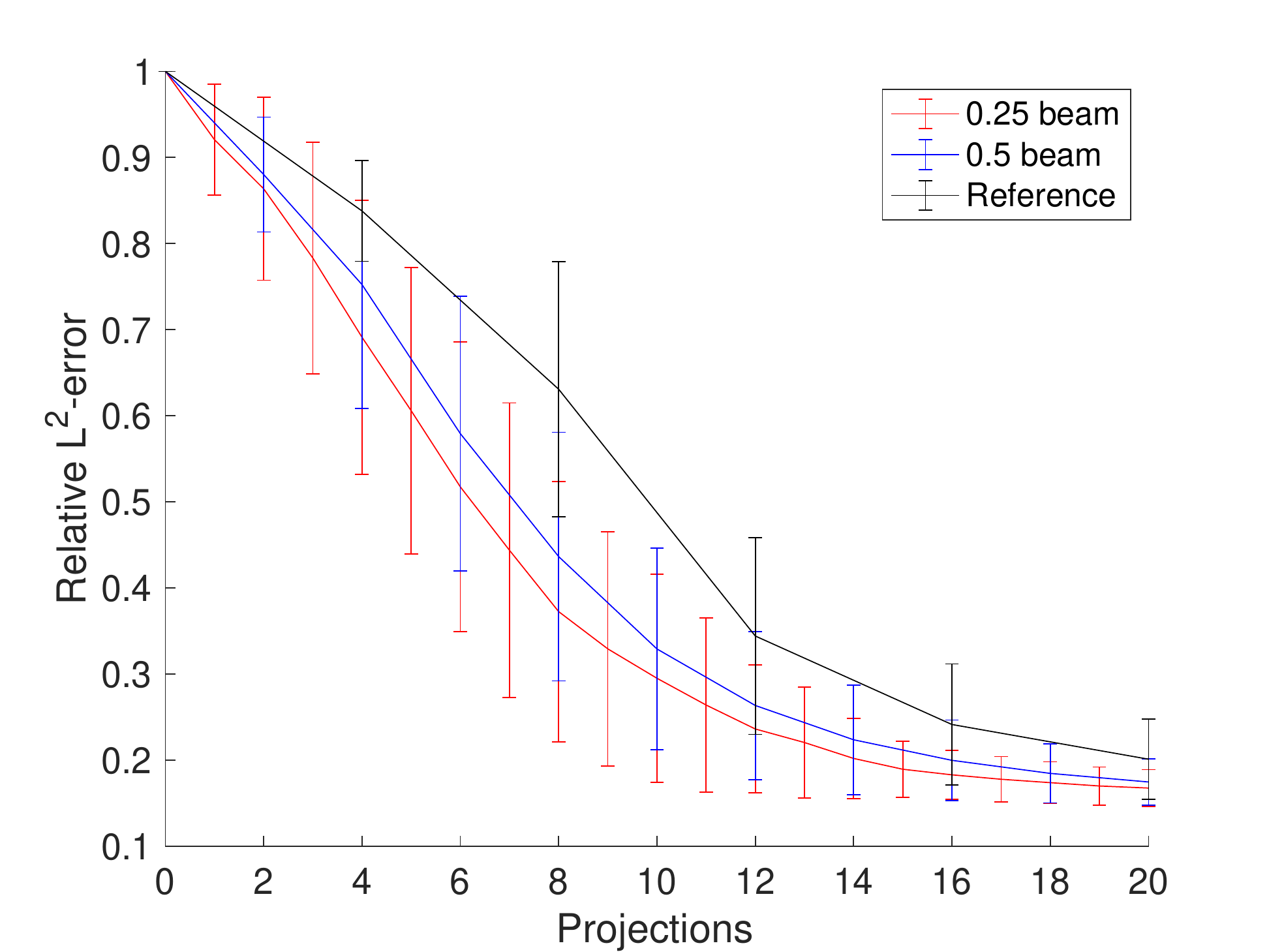}
	\caption{{\sc 2D~Test~2.} Left: Example of a random target composed of  ellipses with randomly chosen shapes, sizes, positions and absorption levels. Right: Mean relative $L^2(D)$ reconstruction errors over $100$ samples for optimized projection geometries and equiangular full-width projections with equivalent radiation doses. The red and blue curves show the errors with optimized projection geometries with beam widths $0.25$ and $0.5$, respectively, whereas the black curve shows the errors for equiangular projections with beam width $1$. The error-bars show the confidence intervals of one standard deviation, and the horizontal axis indicates the number of projections with the narrowest beam width.}
	\label{fig:test2}
\end{figure}

The targets consist of ellipses with constant absorption levels in a homogeneous nonabsorbing background. The number of ellipses is drawn from the uniform distribution over $\{2, 3, 4, 5 \}$, their absorption levels from the uniform distribution over $[0.5,1.5]$ and their centers from the uniform distribution over the disk of radius $0.5$ centered at the midpoint of $D$. Furthermore, the ellipses have (uniformly) random orientations and their semi-major and semi-minor axes are independently drawn from the uniform distribution over $[0.05,0.2]$. In the regions where many ellipses overlap, the absorption level is defined to be the sum of those of the involved ellipses. An example of such a target is shown on the left in Figure \ref{fig:test2}. In particular, note that the ellipses may extend over the domain boundary, which is not in line with the Dirichlet boundary condition for \eqref{eq:diffop} but assures that any considered X-ray may pass through something interesting in a target.

 The discretization of $D$ is the same as in the previous example, that is, the reconstructions are formed on a uniform grid of $n=N^2=10^4$ pixels and a full-width source receiver pair corresponds to $m=51$ individual X-rays (and the $0.25$ and $0.5$ beam widths to $13$ and $26$ X-rays, respectively). However, encouraged by the observations in the previous test, the sequential optimization of the projection geometries is carried out on the sparser grid of $\tilde{n} = \tilde{N}^2 \approx 10^3$ pixels. The total number of considered random targets is $100$ and the noise level is once again set to $\sigma = 10^{-3}$. To make the radiation doses comparable, the algorithm is run for $20$ and $10$ iterations for the beam widths of $0.25$ and $0.5$, respectively, and the corresponding relative $L^2(D)$ reconstruction errors are computed after each iteration. Analogously, the reference reconstructions and the corresponding relative $L^2(D)$ errors are computed for $1$ $2$, $3$, $4$ and $5$ equiangular full-width projections.

 The results, shown on the right in Figure~\ref{fig:test2}, indicate that the error for the optimized projections decreases faster as a function of the radiation dose than that for the reference projections. As in the previous experiment, once enough projection data has been collected, this advantage starts to decrease. Performing the sequential experimental design with a narrower beam seems to be advantageous, presumably because the algorithm can concentrate on retrieving information on certain interesting local details in the target without `wasting radiation'. However, this advantage comes with a fairly significant computational price: in addition to having to run the algorithm for twice as many iterations, the search space is also much wider due to the increased number of possible lateral positions for the source-receiver pair. This poses a problem for our exhaustive optimization routine~\cite{Burger21}. In addition, the overlapping confidence intervals in Figure~\ref{fig:test2} hint that the best approach is target-dependent.

\subsubsection{2D~Test~3: Shepp--Logan phantom}
\begin{figure}
	\centering
	\includegraphics[width = 0.32\columnwidth]{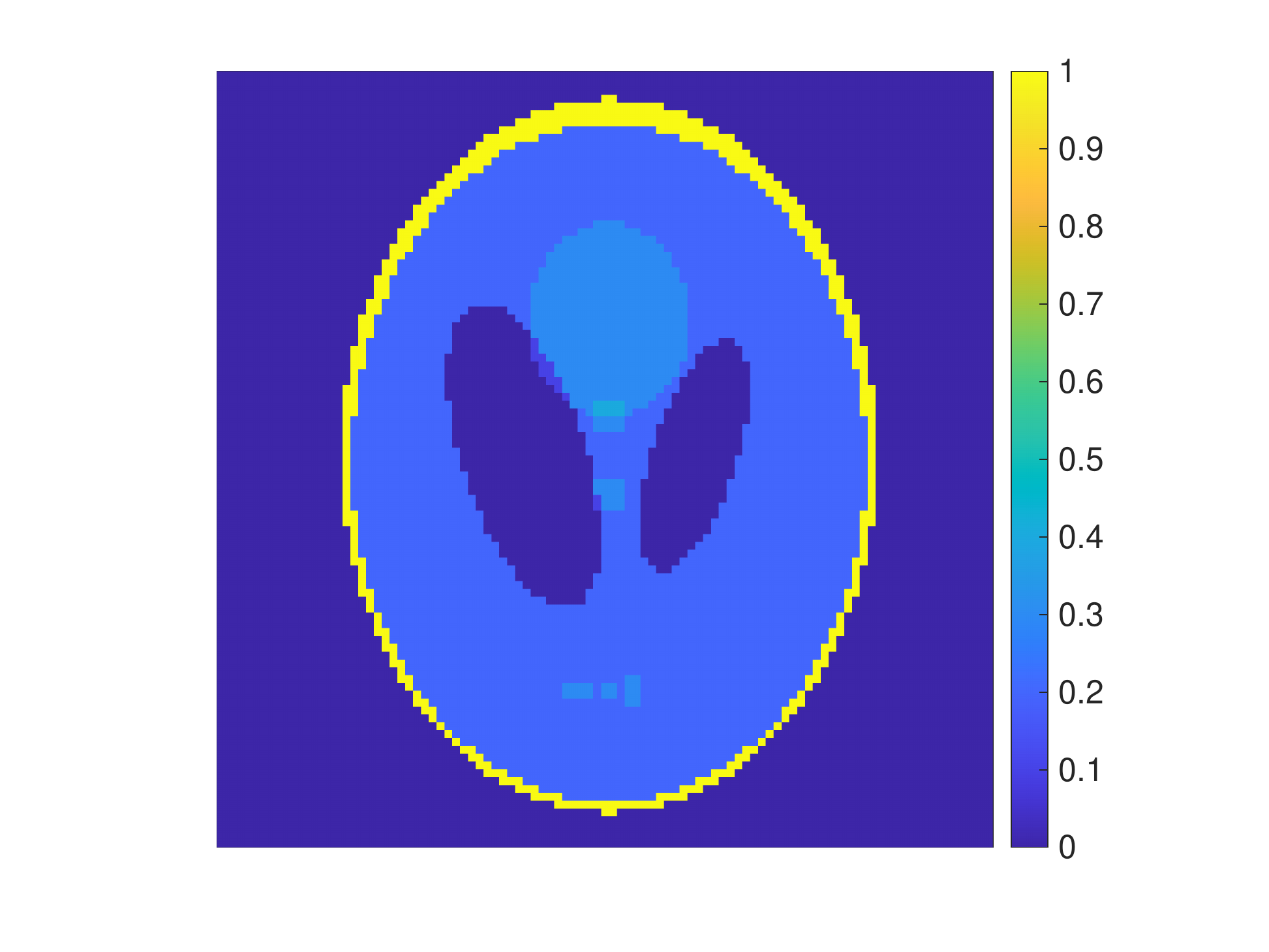}
	\includegraphics[width = 0.40\columnwidth]{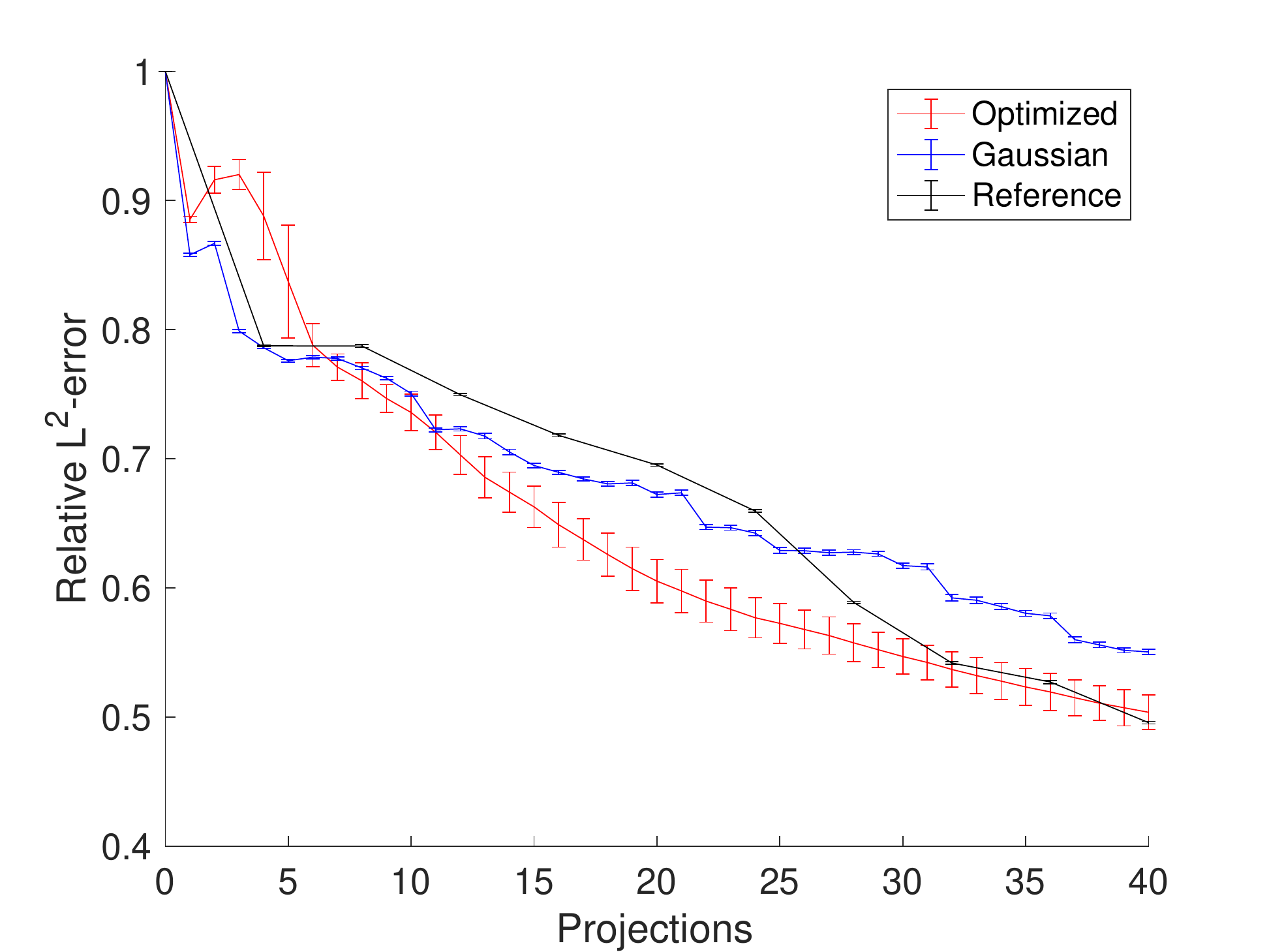} \\[2mm]
	\includegraphics[width = 0.32\columnwidth]{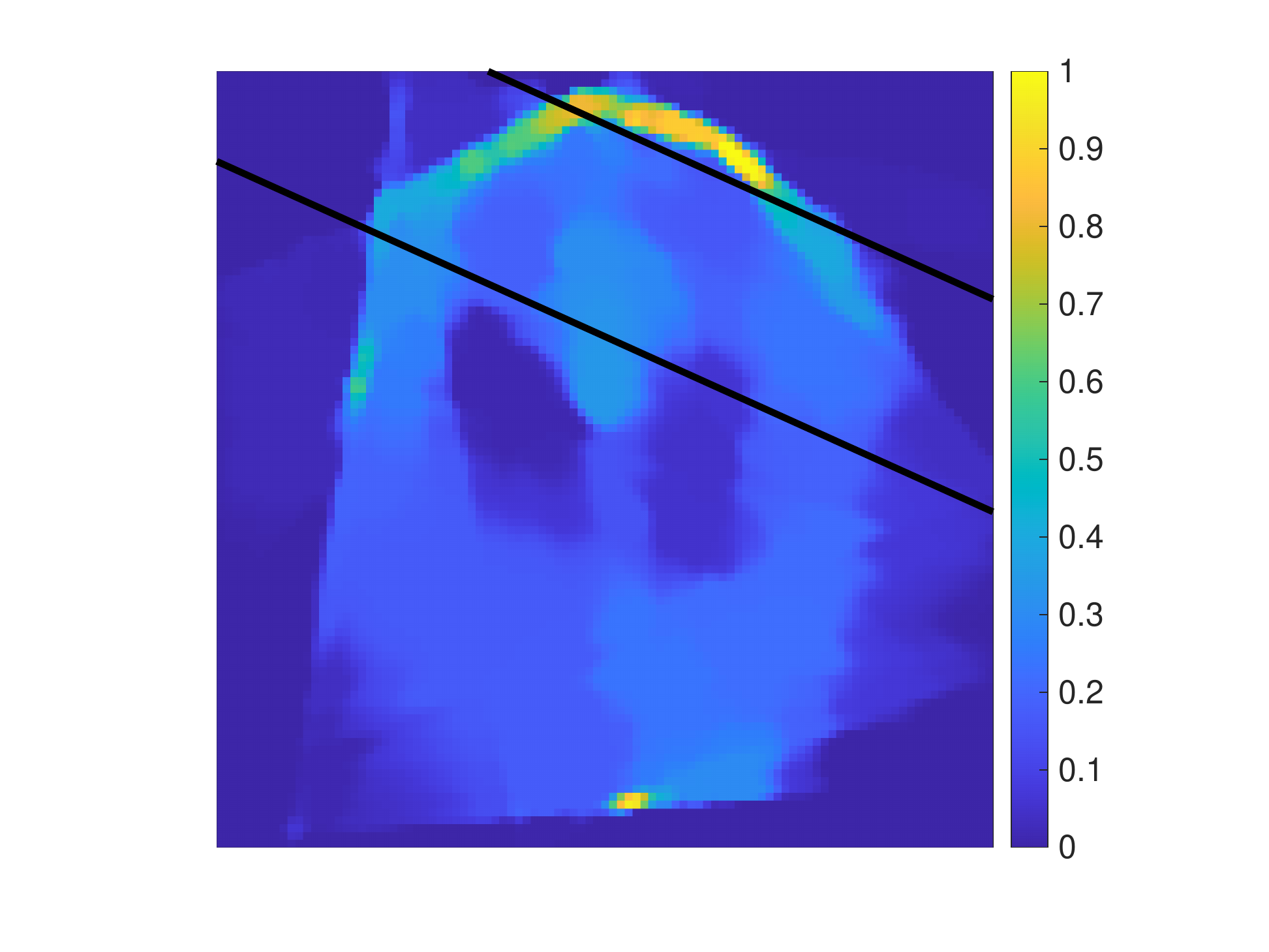}
	\includegraphics[width = 0.32\columnwidth]{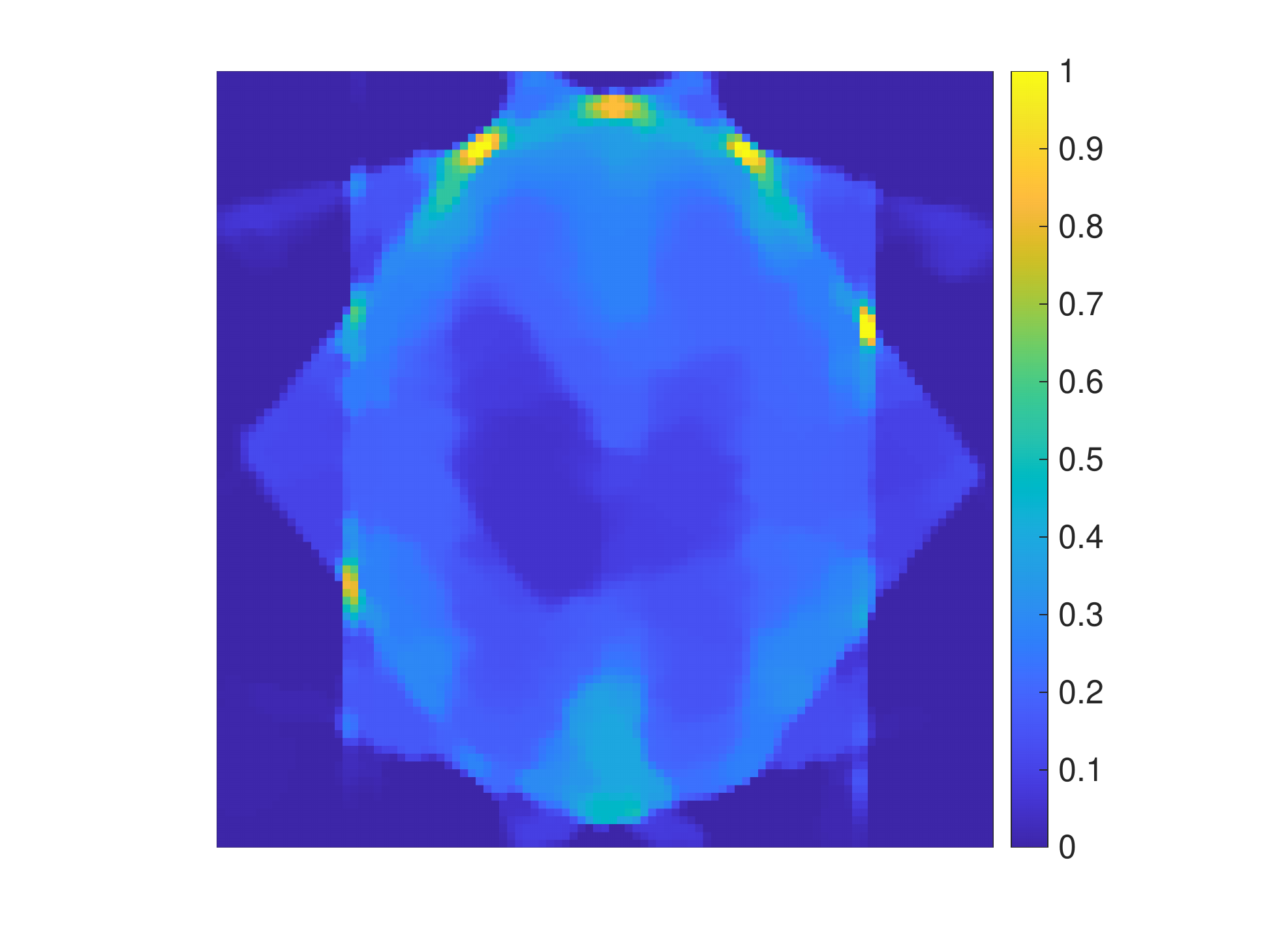}
        \includegraphics[width = 0.32\columnwidth]{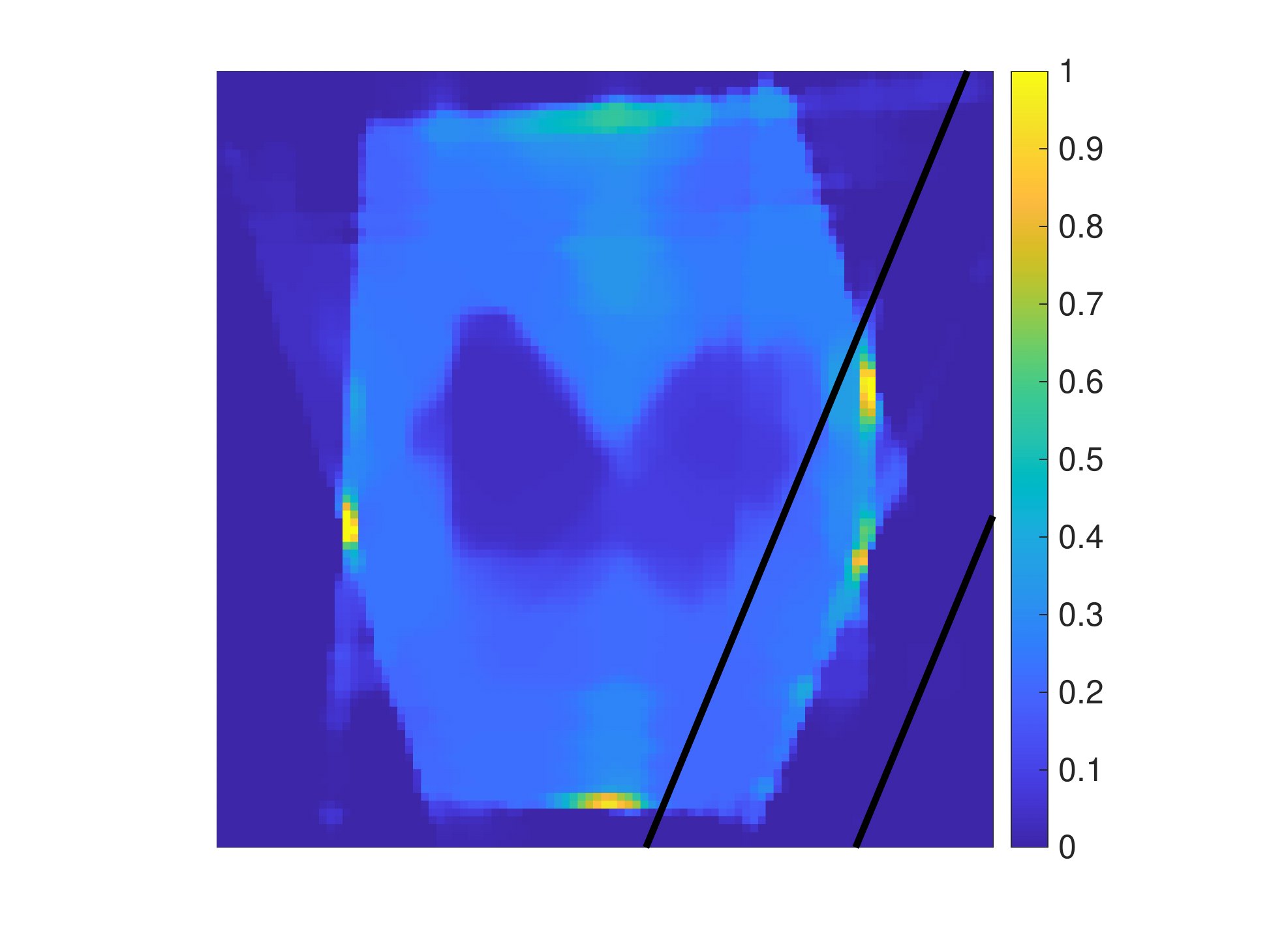}\\[2mm]
        \includegraphics[width = 0.32\columnwidth]{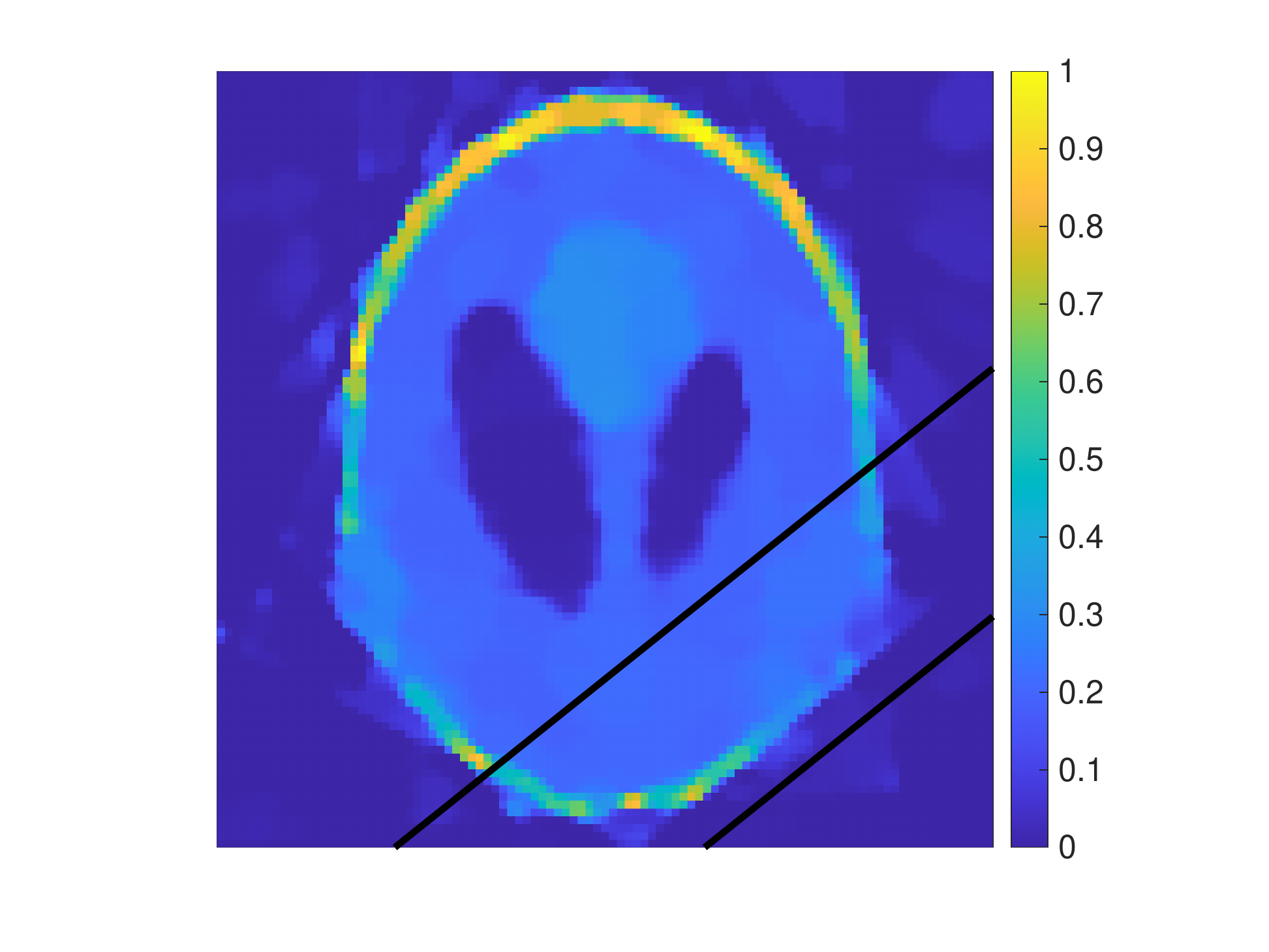}
	\includegraphics[width = 0.32\columnwidth]{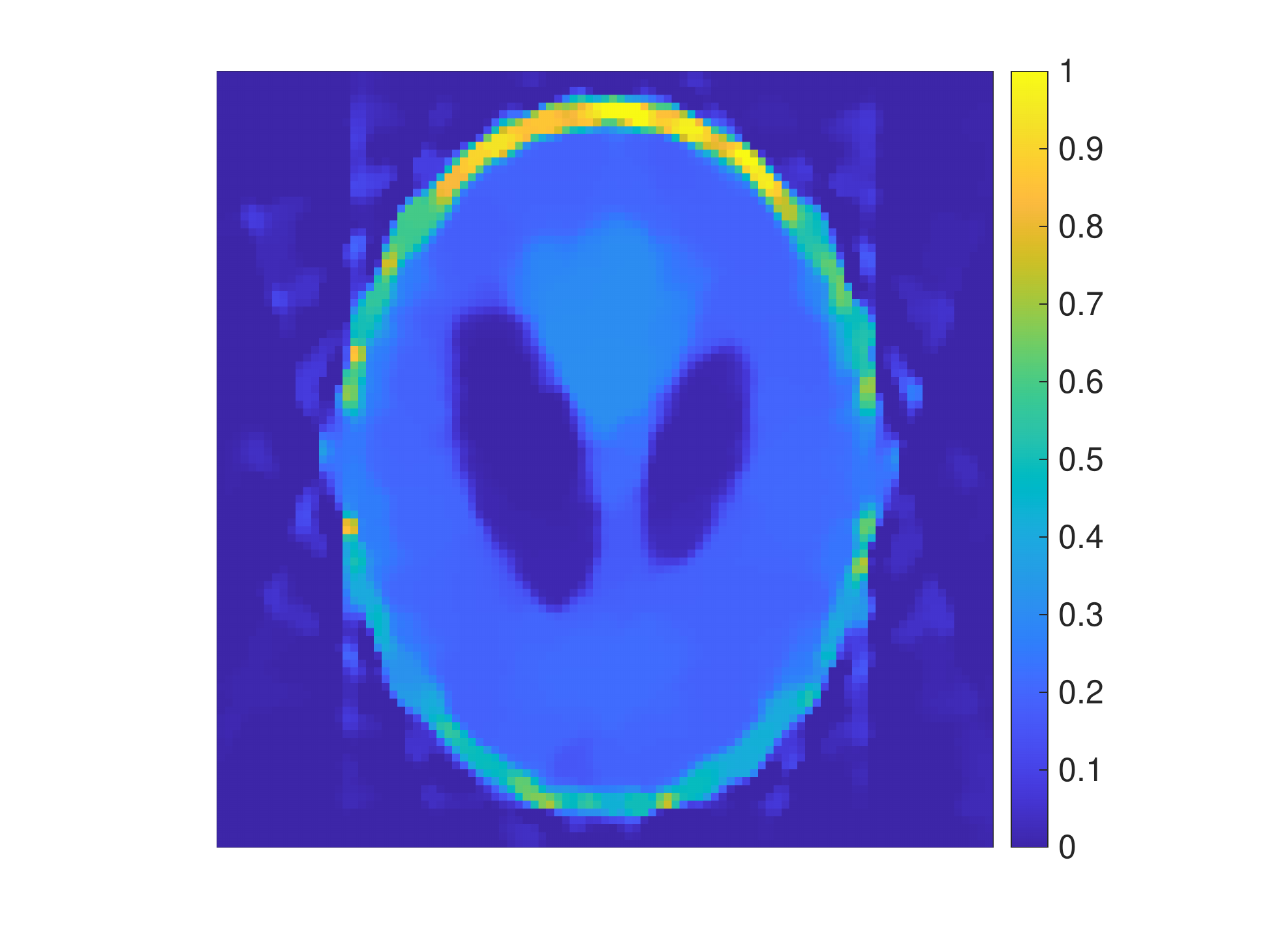}
        \includegraphics[width = 0.32\columnwidth]{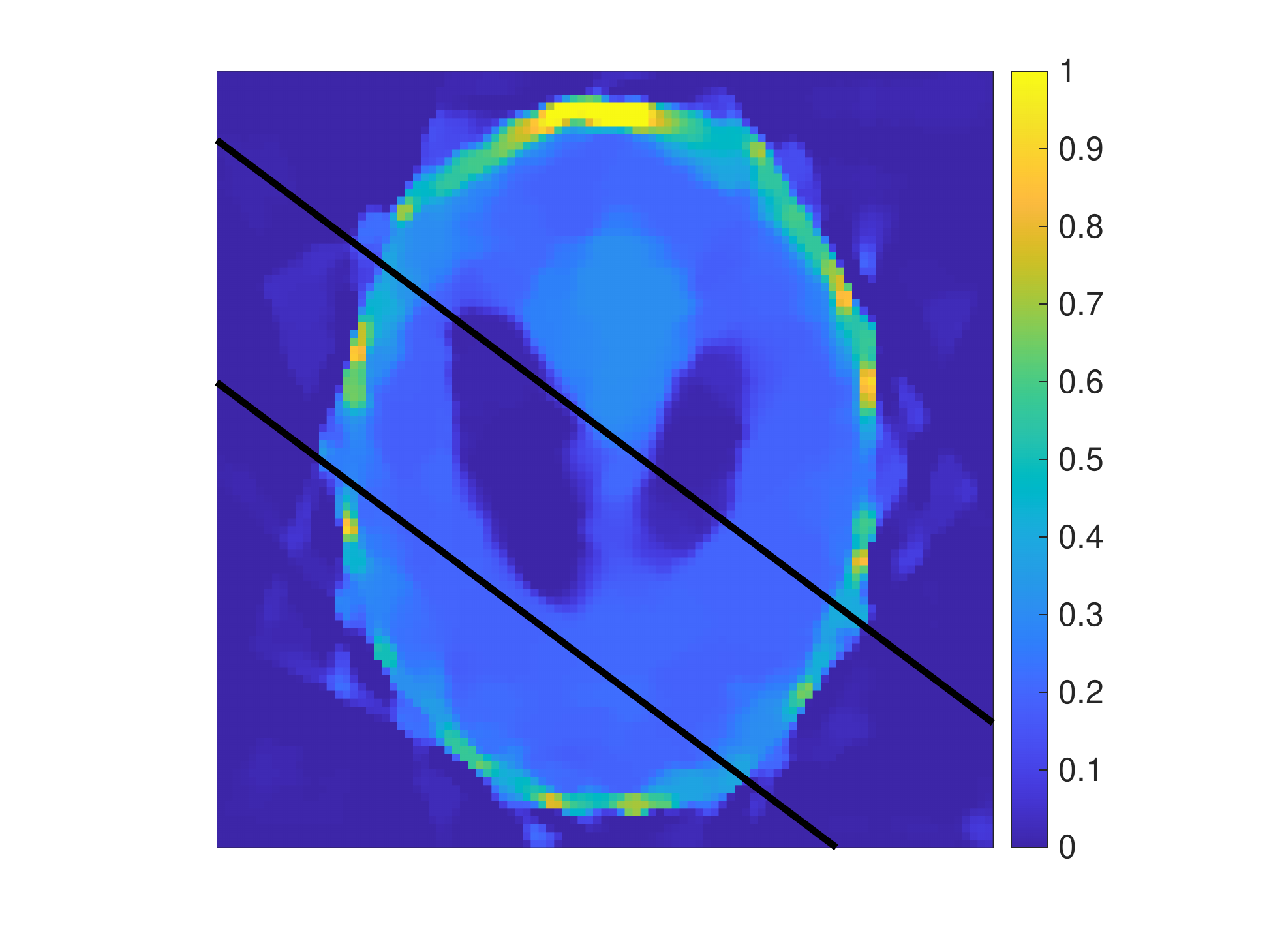}
	\caption{{\sc 2D~Test~3.} Top left: Shepp--Logan phantom. Top right: Mean relative $L^2(D)$ reconstruction errors over $100$ samples of noise realizations for sequentially optimized projection geometries (red: Algorithm~\ref{alg:basic_optimization}, blue: Gaussian prior) and equiangular full-width projections with equivalent radiation doses (black). The error-bars show the confidence intervals of one standard deviation, and the horizontal axis indicates the number of projections with beam width $0.25$. Middle row: Radiation dose corresponding to 5 full-width projections. Bottom row: Radiation dose corresponding to 10 full-width projections. Left column: Reconstructions for projection geometries optimized by Algorithm~\ref{alg:basic_optimization} with the latest projection depicted. Center column: Reconstructions for equiangular full-width projections. Right column: Reconstructions for projection geometries optimized based on a Gaussian prior with the latest projection depicted.}
	\label{fig:test3}
\end{figure}

In our third experiment, the target is the Shepp--Logan phantom shown in the top left image of Figure~\ref{fig:test3}. The main aim is once again to compare the performance of Algorithm~\ref{alg:basic_optimization} with beam width $0.25$ to reconstructions obtained from equiangular full-width reference projections. However, we also consider using in Algorithm~\ref{alg:basic_optimization} sequentially optimized quarter-width projections corresponding to a Gaussian prior with a covariance matrix of the form
\begin{equation}
  \label{eq:Gaussian}
 (\Gamma_{\rm prior})_{i,j} = \eta^2 \exp \left(-\frac{| x_i - x_j |^2}{2\ell^2} \right).
 \end{equation}
Here $| \cdot |$ denotes the Euclidean norm, $\ell>0$ is the so-called correlation length, $\eta>0$ is the pixelwise standard deviation, and $x_i$ denotes the center of the $i$th pixel. Under such a prior, the sequentially optimized projections do not depend on the measurements or the prior mean, and they can thus be computed in advance based on merely the covariance matrix~\eqref{eq:Gaussian} and the known structure of the additive Gaussian noise process; see~\cite{Burger21} for more details. When employing a Gaussian prior with the covariance structure~\eqref{eq:Gaussian}, we thus use in Algorithm~\ref{alg:basic_optimization} precomputed sequentially optimized design variables instead of determining the projection geometries adaptively as a part of the algorithm itself. However, the lagged diffusivity iteration is still employed in the computation of the reconstructions, as indicated by the interior loop of Algorithm~\ref{alg:basic_optimization}.

We choose $\eta = 0.2$ and $\ell=0.1$ in \eqref{eq:Gaussian}; the former is close to the pixelwise standard deviation of the Shepp--Logan phantom, whereas the latter simply seems to be in a relatively good agreement with the sizes of the areas with constant absorption in the top left image of Figure~\ref{fig:test3}. All other parameters are the same as in the previous experiment. In particular, the optimization steps of Algorithm~\ref{alg:basic_optimization} are once again carried out on a sparser grid with $\tilde{N}^2 \approx 10^3$ pixels, and this same sparse discretization is also used for deducing the sequentially optimized projection geometries corresponding to the Gaussian prior with the covariance matrix \eqref{eq:Gaussian}. The test is run $100$ times to examine how the measurement noise affects the reconstruction quality, mainly via changes in the adaptive optimal designs produced by Algorithm~\ref{alg:basic_optimization}.

The top right image in Figure \ref{fig:test3} shows the (mean) relative $L^2(D)$ reconstruction errors up to $40$ and $10$ projections for the two types of sequentially optimized geometries and the equiangular reference geometries, respectively. After a radiation dose that is equivalent to one full-width projection, both sets of reference projections, i.e.~the full-width equiangular one and the four quarter-width ones based on the Gaussian prior, correspond to lower $L^2(D)$ errors than the reconstruction produced by Algorithm~\ref{alg:basic_optimization}. This is likely due to the Shepp--Logan phantom covering most of the domain, which makes an initial full-width projection or four rather randomly distributed, non-adaptively chosen quarter-width projections sensible approaches. During the following 36 rounds of Algorithm~\ref{alg:basic_optimization}, the edge-promoting sequentially optimal design first shows a clear advantage over the full-width reference reconstructions, but the advantage diminishes after enough projection data has been collected. On the other hand, the reference quarter-width projections based on the Gaussian prior perform almost as well as Algorithm~\ref{alg:basic_optimization} until about 10 projections, but subsequently the adaptive approach of Algorithm~\ref{alg:basic_optimization} leads to clearly superior results. It is also interesting to notice that the equiangular full-width heuristic starts to outperform the sequentially optimized projections corresponding to the Gaussian prior at about 25 quarter-width projections.

The final reconstructions after 40 sequentially optimized and with 10 full-width equiangular projections, as well as those after only 20 optimized and 5 equiangular projections, are presented on the bottom and middle rows of Figure~\ref{fig:test3}. The reference reconstructions corresponding to the full-width projections in the middle column contain characteristic streaking artifacts of sparse-angle X-ray tomography, evenly spread around the target; this effect is particularly emphasized in the case of five full-width projections. For the 20 adaptively optimized projections in the left-hand column, some sections of the `head boundary' are reconstructed more accurately than in the corresponding reference reconstructions, and there is arguably also less blurring in the interior of the phantom. However, with only 20 projections Algorithm~\ref{alg:basic_optimization} leads to a bad reconstruction of the bottom half of the phantom as the optimized projections have not yet covered that region comprehensively. This exemplifies an obvious flaw in our approach: even if the sequentially chosen projection geometries were locally optimal, their combination is no longer optimal after several rounds, and there is no guarantee that this nonoptimality could not be severe if a high number of projection geometries is considered. The final reconstruction produced by Algorithm~\ref{alg:basic_optimization} after 40 quarter-width projections and the one corresponding to ten equiangular full-width projections shown on the bottom row of Figure~\ref{fig:test3} are comparable in quality, whereas the one corresponding to the 40 quarter-width projections sequentially optimized based on the Gaussian prior is arguably somewhat inferior. These observations are inline with the information in the convergence plot of the top right image in Figure~\ref{fig:test3}.

\subsection{Three-dimensional cone beam tomography}
\begin{figure}
	\centering
	\tdplotsetmaincoords{70}{150}
\begin{tikzpicture}[scale=3,tdplot_main_coords]
    \coordinate (O) at (0,0,0);
    \tdplotsetcoord{P}{1.732}{53.776}{45}

    \draw[thick,->] (0,0,0) -- (1.5,0,0) node[anchor=north east]{$x$};
    \draw[thick,->] (0,0,0) -- (0,1.2,0) node[anchor=north west]{$y$};
    \draw[thick,->] (0,0,0) -- (0,0,1.3) node[anchor=south]{$z$};

    \draw[dashed, color=blue] (O) -- (1,0,0);
    \draw[dashed, color=red] (O) -- (Px);
    \draw[dashed, color=red] (O) -- (Py);
    \draw[dashed, color=red] (O) -- (Pz);
    \draw[dashed, color=red] (Px) -- (Pxy);
    \draw[dashed, color=red] (Py) -- (Pxy);
    \draw[dashed, color=red] (Px) -- (Pxz);
    \draw[dashed, color=red] (Pz) -- (Pxz);
    \draw[dashed, color=red] (Py) -- (Pyz);
    \draw[dashed, color=red] (Pz) -- (Pyz);
    \draw[dashed, color=red] (Pxy) -- (P);
    \draw[dashed, color=red] (Pxz) -- (P);
    \draw[dashed, color=red] (Pyz) -- (P);

    \node at (-1.5, 0.5, 0.5) {\textbullet};
    \node at (-1.5, 0.5, 0.5) [above] {Source};

\node at (1.1711409395973154,0.9697986577181208,1.0734623443633282) {\textbullet};
\node at (1.2813325221960805,1.0469327655372564,0.8006393848790938) {\textbullet};
\node at (1.3192319205190404,1.0734623443633282,0.5000000000000001) {\textbullet};
\node at (1.2813325221960805,1.0469327655372564,0.19936061512090675) {\textbullet};
\node at (1.1711409395973156,0.9697986577181208,-0.0734623443633281) {\textbullet};
\node at (1.2813325221960805,0.7462933806581628,1.0734623443633282) {\textbullet};
\node at (1.4096159602595202,0.7867311721816641,0.8006393848790938) {\textbullet};
\node at (1.4537378886567947,0.8006393848790936,0.5000000000000001) {\textbullet};
\node at (1.4096159602595204,0.7867311721816642,0.19936061512090675) {\textbullet};
\node at (1.2813325221960807,0.7462933806581629,-0.0734623443633281) {\textbullet};
\node at (1.3192319205190404,0.5,1.0734623443633282) {\textbullet};
\node at (1.4537378886567947,0.5,0.8006393848790938) {\textbullet};
\node at (1.5,0.5,0.5000000000000001) {\textbullet};
\node at (1.4537378886567947,0.5,0.19936061512090675) {\textbullet};
\node at (1.3192319205190406,0.5,-0.0734623443633281) {\textbullet};
\node at (1.2813325221960805,0.25370661934183714,1.0734623443633282) {\textbullet};
\node at (1.4096159602595202,0.21326882781833578,0.8006393848790938) {\textbullet};
\node at (1.4537378886567947,0.19936061512090636,0.5000000000000001) {\textbullet};
\node at (1.4096159602595204,0.21326882781833578,0.19936061512090675) {\textbullet};
\node at (1.2813325221960807,0.25370661934183714,-0.0734623443633281) {\textbullet};
\node at (1.1711409395973154,0.030201342281879262,1.0734623443633282) {\textbullet};
\node at (1.2813325221960805,-0.04693276553725645,0.8006393848790938) {\textbullet};
\node at (1.3192319205190404,-0.07346234436332832,0.5000000000000001) {\textbullet};
\node at (1.2813325221960805,-0.04693276553725645,0.19936061512090675) {\textbullet};
\node at (1.1711409395973156,0.03020134228187915,-0.0734623443633281) {\textbullet};
\draw (1.1711409395973154,0.9697986577181208,1.0734623443633282) -- (1.2813325221960805,0.7462933806581628,1.0734623443633282);
\draw (1.2813325221960805,1.0469327655372564,0.8006393848790938) -- (1.4096159602595202,0.7867311721816641,0.8006393848790938);
\draw (1.3192319205190404,1.0734623443633282,0.5000000000000001) -- (1.4537378886567947,0.8006393848790936,0.5000000000000001);
\draw (1.2813325221960805,1.0469327655372564,0.19936061512090675) -- (1.4096159602595204,0.7867311721816642,0.19936061512090675);
\draw (1.1711409395973156,0.9697986577181208,-0.0734623443633281) -- (1.2813325221960807,0.7462933806581629,-0.0734623443633281);
\draw (1.2813325221960805,0.7462933806581628,1.0734623443633282) -- (1.3192319205190404,0.5,1.0734623443633282);
\draw (1.4096159602595202,0.7867311721816641,0.8006393848790938) -- (1.4537378886567947,0.5,0.8006393848790938);
\draw (1.4537378886567947,0.8006393848790936,0.5000000000000001) -- (1.5,0.5,0.5000000000000001);
\draw (1.4096159602595204,0.7867311721816642,0.19936061512090675) -- (1.4537378886567947,0.5,0.19936061512090675);
\draw (1.2813325221960807,0.7462933806581629,-0.0734623443633281) -- (1.3192319205190406,0.5,-0.0734623443633281);
\draw (1.3192319205190404,0.5,1.0734623443633282) -- (1.2813325221960805,0.25370661934183714,1.0734623443633282);
\draw (1.4537378886567947,0.5,0.8006393848790938) -- (1.4096159602595202,0.21326882781833578,0.8006393848790938);
\draw (1.5,0.5,0.5000000000000001) -- (1.4537378886567947,0.19936061512090636,0.5000000000000001);
\draw (1.4537378886567947,0.5,0.19936061512090675) -- (1.4096159602595204,0.21326882781833578,0.19936061512090675);
\draw (1.3192319205190406,0.5,-0.0734623443633281) -- (1.2813325221960807,0.25370661934183714,-0.0734623443633281);
\draw (1.2813325221960805,0.25370661934183714,1.0734623443633282) -- (1.1711409395973154,0.030201342281879262,1.0734623443633282);
\draw (1.4096159602595202,0.21326882781833578,0.8006393848790938) -- (1.2813325221960805,-0.04693276553725645,0.8006393848790938);
\draw (1.4537378886567947,0.19936061512090636,0.5000000000000001) -- (1.3192319205190404,-0.07346234436332832,0.5000000000000001);
\draw (1.4096159602595204,0.21326882781833578,0.19936061512090675) -- (1.2813325221960805,-0.04693276553725645,0.19936061512090675);
\draw (1.2813325221960807,0.25370661934183714,-0.0734623443633281) -- (1.1711409395973156,0.03020134228187915,-0.0734623443633281);
\draw (1.1711409395973154,0.9697986577181208,1.0734623443633282) -- (1.2813325221960805,1.0469327655372564,0.8006393848790938);
\draw (1.2813325221960805,1.0469327655372564,0.8006393848790938) -- (1.3192319205190404,1.0734623443633282,0.5000000000000001);
\draw (1.3192319205190404,1.0734623443633282,0.5000000000000001) -- (1.2813325221960805,1.0469327655372564,0.19936061512090675);
\draw (1.2813325221960805,1.0469327655372564,0.19936061512090675) -- (1.1711409395973156,0.9697986577181208,-0.0734623443633281);
\draw (1.2813325221960805,0.7462933806581628,1.0734623443633282) -- (1.4096159602595202,0.7867311721816641,0.8006393848790938);
\draw (1.4096159602595202,0.7867311721816641,0.8006393848790938) -- (1.4537378886567947,0.8006393848790936,0.5000000000000001);
\draw (1.4537378886567947,0.8006393848790936,0.5000000000000001) -- (1.4096159602595204,0.7867311721816642,0.19936061512090675);
\draw (1.4096159602595204,0.7867311721816642,0.19936061512090675) -- (1.2813325221960807,0.7462933806581629,-0.0734623443633281);
\draw (1.3192319205190404,0.5,1.0734623443633282) -- (1.4537378886567947,0.5,0.8006393848790938);
\draw (1.4537378886567947,0.5,0.8006393848790938) -- (1.5,0.5,0.5000000000000001);
\draw (1.5,0.5,0.5000000000000001) -- (1.4537378886567947,0.5,0.19936061512090675);
\draw (1.4537378886567947,0.5,0.19936061512090675) -- (1.3192319205190406,0.5,-0.0734623443633281);
\draw (1.2813325221960805,0.25370661934183714,1.0734623443633282) -- (1.4096159602595202,0.21326882781833578,0.8006393848790938);
\draw (1.4096159602595202,0.21326882781833578,0.8006393848790938) -- (1.4537378886567947,0.19936061512090636,0.5000000000000001);
\draw (1.4537378886567947,0.19936061512090636,0.5000000000000001) -- (1.4096159602595204,0.21326882781833578,0.19936061512090675);
\draw (1.4096159602595204,0.21326882781833578,0.19936061512090675) -- (1.2813325221960807,0.25370661934183714,-0.0734623443633281);
\draw (1.1711409395973154,0.030201342281879262,1.0734623443633282) -- (1.2813325221960805,-0.04693276553725645,0.8006393848790938);
\draw (1.2813325221960805,-0.04693276553725645,0.8006393848790938) -- (1.3192319205190404,-0.07346234436332832,0.5000000000000001);
\draw (1.3192319205190404,-0.07346234436332832,0.5000000000000001) -- (1.2813325221960805,-0.04693276553725645,0.19936061512090675);
\draw (1.2813325221960805,-0.04693276553725645,0.19936061512090675) -- (1.1711409395973156,0.03020134228187915,-0.0734623443633281);
\draw[dashed, color=gray] (-1.5, 0.5, 0.5) --(1.1711409395973154,0.9697986577181208,1.0734623443633282);
\draw[dashed, color=gray] (-1.5, 0.5, 0.5) --(1.1711409395973156,0.9697986577181208,-0.0734623443633281);
\draw[dashed, color=gray] (-1.5, 0.5, 0.5) --(1.1711409395973154,0.030201342281879262,1.0734623443633282);
\draw[dashed, color=gray] (-1.5, 0.5, 0.5) --(1.1711409395973156,0.03020134228187915,-0.0734623443633281);
\end{tikzpicture}
	\caption{Measurement setup of three-dimensional cone beam tomography.}
	\label{fig:3dgeom}
\end{figure}
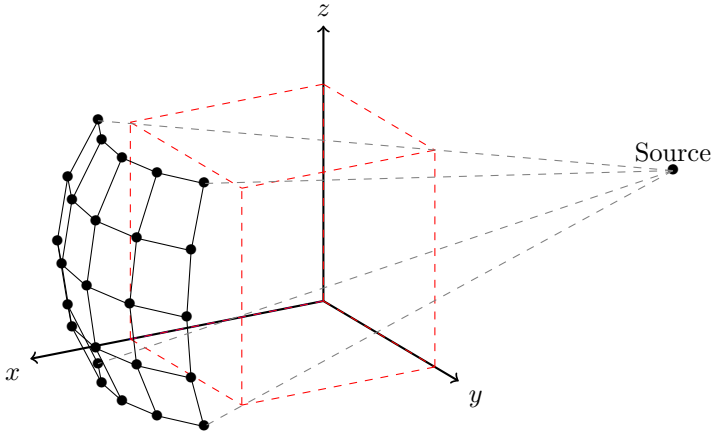

In three dimensions, the unknown absorption distribution is located in the unit cube $D = [0,1]^3$ that is discretized into a uniform grid of $n = N^3$ voxels. We consider cone beam tomography, where a point-like source at $s\in \R^3$ sends X-rays to a two-dimensional receiver patch that occupies a `square' solid angle of the form $[\theta + \delta, \theta - \delta] \times [\phi + \delta, \phi - \delta]$ if the origin is transferred to $s$ without affecting the orientation of the coordinate axes; see Figure \ref{fig:3dgeom}. Here $\theta$ and $\phi$ denote the central polar and azimuthal angles of the detector, respectively. When considering full-aperture projections, the imaging system is always aligned so that the line between the source and the midpoint of the receiver passes through the center of the cube $D$. The receiver is discretized into a rectangular grid of $m = M^2$ detectors with respect to its polar and azimuthal angles in the coordinate system centered at $s$. To summarize, a single full-aperture projection geometry is defined by the central spherical angles of the detector $(\theta, \phi)$ with respect to the source $s$ (or the center of $D$), the corresponding opening angle $\delta$, the distance $d$ from the source to the center of $D$, and the number of pixels per edge $M$ in the detector. Observe that the distance between the source and the detector does not play a role as long as the two are on opposite sides of $D$.

After assigning (fixed) values for $d$, $\delta$ and $M$, a set of full-aperture projection geometries to be used in the exhaustive optimization algorithm of \cite{Burger21} is defined by choosing the corresponding central spherical angles $(\theta_j, \phi_j)$. Unlike in the two dimensional examples with parallel beam tomography, the projections are not symmetric with respect to reflections about the center of the object, and thus one cannot only focus on projections from one side of the object,~i.e.,~one cannot exclude some projection directions as redundant by a symmetry argument. To simulate movement of a smaller detector in the lateral direction, it is possible to only consider some subset of detectors in a full-aperture receiver.

\subsubsection{3D~Test with simple geometric shapes}
\begin{figure}
 \centering
 \includegraphics[width = 0.49\columnwidth]{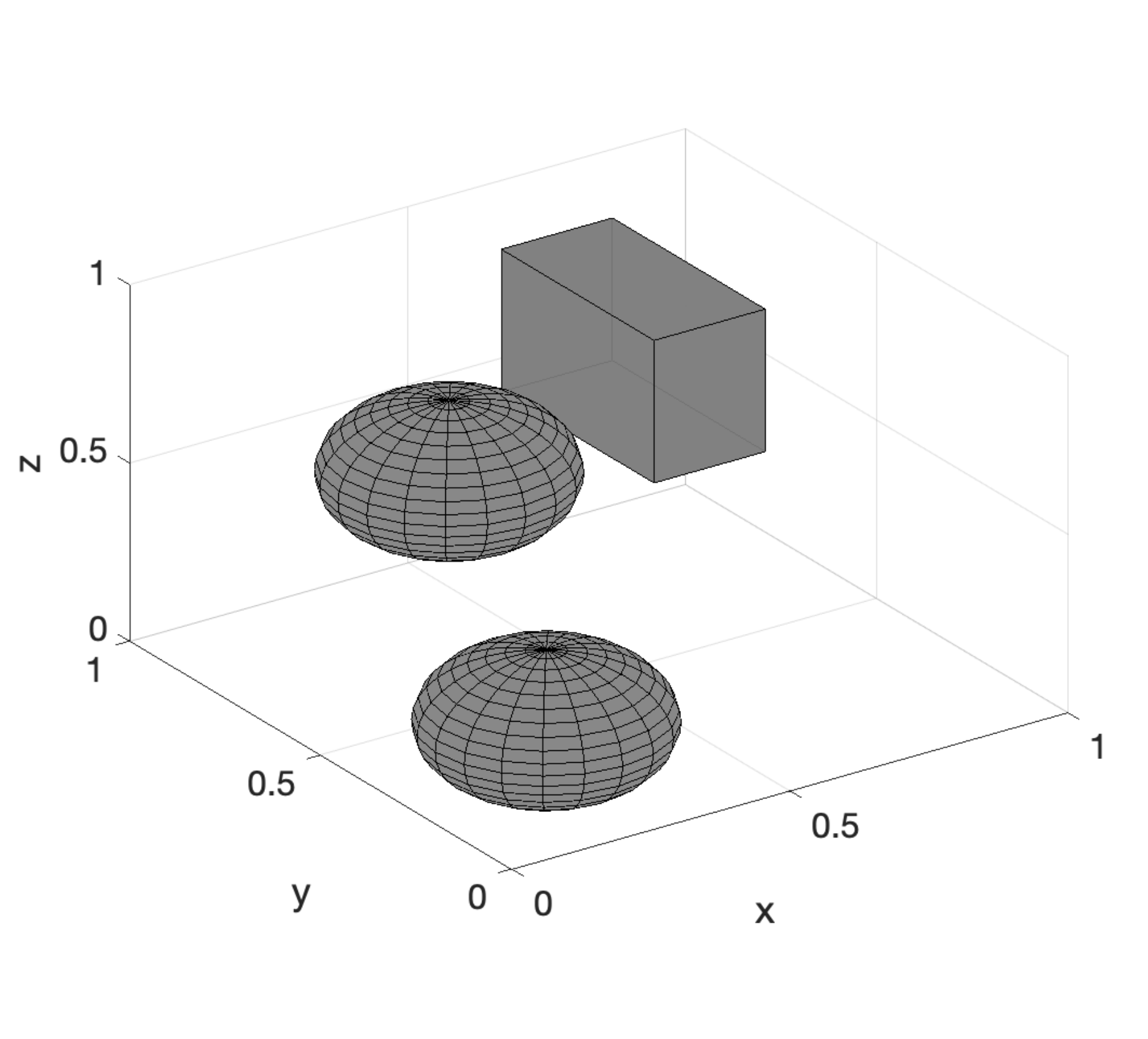}
 \includegraphics[width = 0.49\columnwidth]{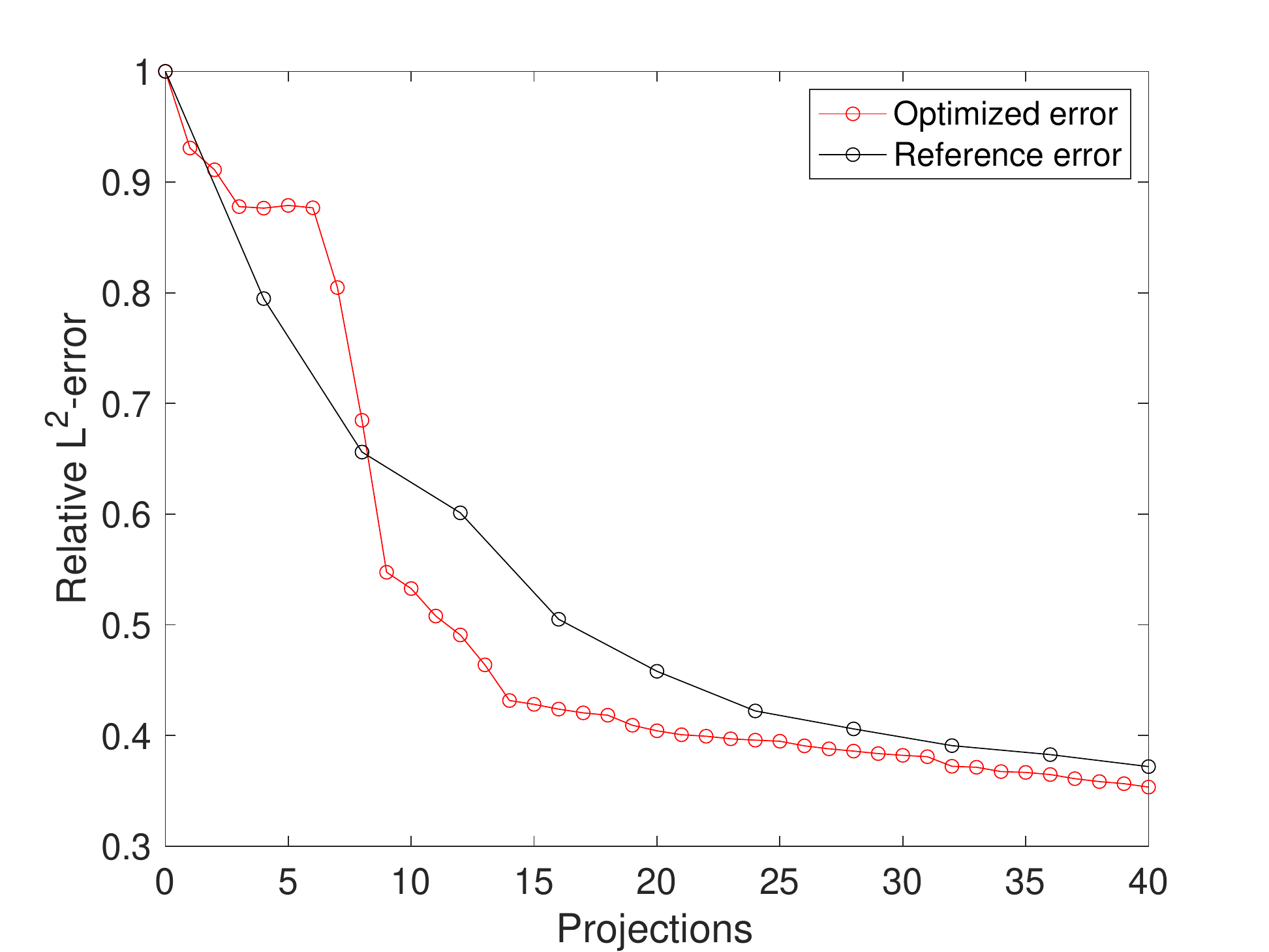}
 \caption{{\sc 3D~Test.} Left: Target. Right: Relative $L^2(D)$ reconstruction errors for optimized quarter-aperture projection geometries and `equally spaced' full-aperture projections with equivalent radiation doses. The red curve depicts the errors for optimized projection geometries whereas the black curve shows those for the equally spaced projections. The horizontal axis indicates the number of quarter-aperture projections.}
 \label{fig:3Dtest1}
\end{figure}

Analogously to the first two-dimensional experiment, our three-dimensional example only aims at demonstrating the basic operation of the algorithm. The target shown on the left in Figure~\ref{fig:3Dtest1} consists of two balls with radius $0.2$ centered at $(0.2,0.2,0.2)$ and $(0.3,0.6,0.6)$, respectively, and a rectangular cuboid $[0.6, 0.8] \times [0.5,0.9] \times [0.5,0.9]$ in a homogeneous background with vanishing absorption.
The common constant absorption level of the balls is~$1$, and that of the cuboid is $2$. The target $D$ is discretized into a grid with $N=50$ voxels per edge, i.e.~a total of $1.25\cdot 10^5$ unknowns. For the optimization step of Algorithm~\ref{alg:basic_optimization}, we interpolate once again onto a {\em significantly} sparser grid with $20^3 = 8000$ pixels to speed up the computations. The noise level is chosen to be $\sigma = 2 \cdot 10^{-3}$, the opening angle of the projection cones is $\delta = 0.24$ radians, and the distance from the source to the center of $D$ is set to~$2.5$.

To define the set of (central) projection angles used in determining the search space for the exhaustive optimization algorithm from~\cite{Burger21}, we introduce $60$ evenly spaced azimuthal angles $\phi_i$ over the interval $[0, 2 \pi]$ and three polar angles $-\pi/4, 0, \pi/4$, with the zero polar angle associated to directions parallel to the xy-plane. The total set of projection directions is then $[\phi_1, \ldots,\phi_{30}] \times [-\pi/4, 0, \pi/4]$. The detector is split into four quadrants, each with $10 \times 10$ detectors, to allow four quarter-aperture projection geometries for each projection direction. This construction results in a total number of  $4 \times 3 \times 60 = 720$ available projection geometries for the exhaustive algorithm from \cite{Burger21}. In particular, note that the set of possible projection directions is both sparse and limited in the polar direction, which has a certain effect on the achievable reconstruction quality~\cite{Quinto93}.

Algorithm~\ref{alg:basic_optimization} is run for a total of 40 rounds. For reference, we once again also consider reconstructions corresponding to `equally spaced'
full-aperture projections of equivalent radiation dose. Unlike in two-dimensions, there is no obvious methodology for choosing the directions for these reference projections: (i) there exist now fundamental way of uniformly sampling the available 180 directions and (ii) it is obvious that projections from opposite directions contain similar, yet not exactly the same information. Our heuristic for choosing the directions of the full-aperture projections is including in the computation of the reference reconstructions one by one more projection directions from the sequence: $(0,0)$, $(0, 2\pi/3)$, $(0, 4\pi/3)$, $(\pi/4,\pi)$, $(\pi/4, 0)$, $(-\pi/4, \pi/2)$, $(-\pi/4, 3\pi/2)$, $(0, \pi/6)$, $(0, 3\pi/2)$, $(0, 5\pi/6)$. In particular, note that this construction does not even aim at globally optimal reference directions, as are arguably the equiangular directions in two dimensions, but the selection of the reference projection geometries is also sequential in the sense that all previously used projections are also included in the subsequent projection sets of higher cardinality.

The relative $L^2(D)$ reconstruction errors for equivalent radiation doses are shown on the right in Figure~\ref{fig:3Dtest1}. For the optimized quarter-aperture projections, the reconstruction error initially starts to decrease, before plateauing for iterations 3-6. At that point, the reconstruction error for the reference projections decreases faster, with the quality of the reference reconstructions being better for radiation doses equivalent to 3-8 quarter-aperture projections. However, between 7 and 10 iterations of Algorithm~\ref{alg:basic_optimization}, the reconstruction error for the optimized quarter-aperture projections drops rapidly below the reference curve and stays there all the way until the limit of 40 quarter-aperture projections is reached. As in the two-dimensional experiments, once enough data has been collected the optimized quarter-aperture projections and the reference full-width projections result in roughly the same reconstruction errors for equivalent radiation doses.

\begin{figure}
	\centering
	\includegraphics[width = 0.95\columnwidth]{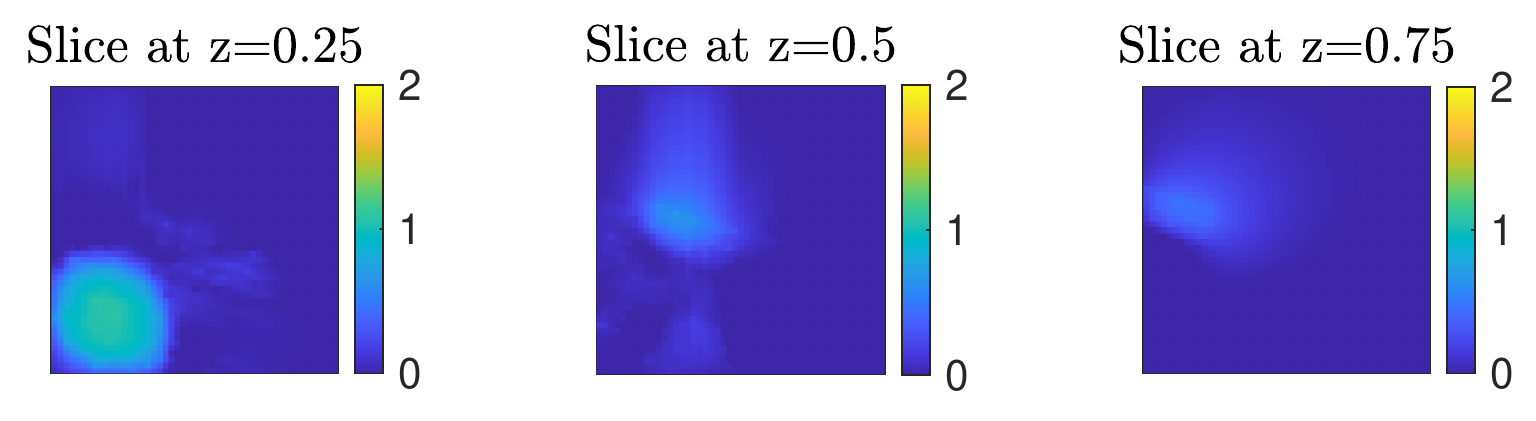} \\[4mm]
	\includegraphics[width = 0.95\columnwidth]{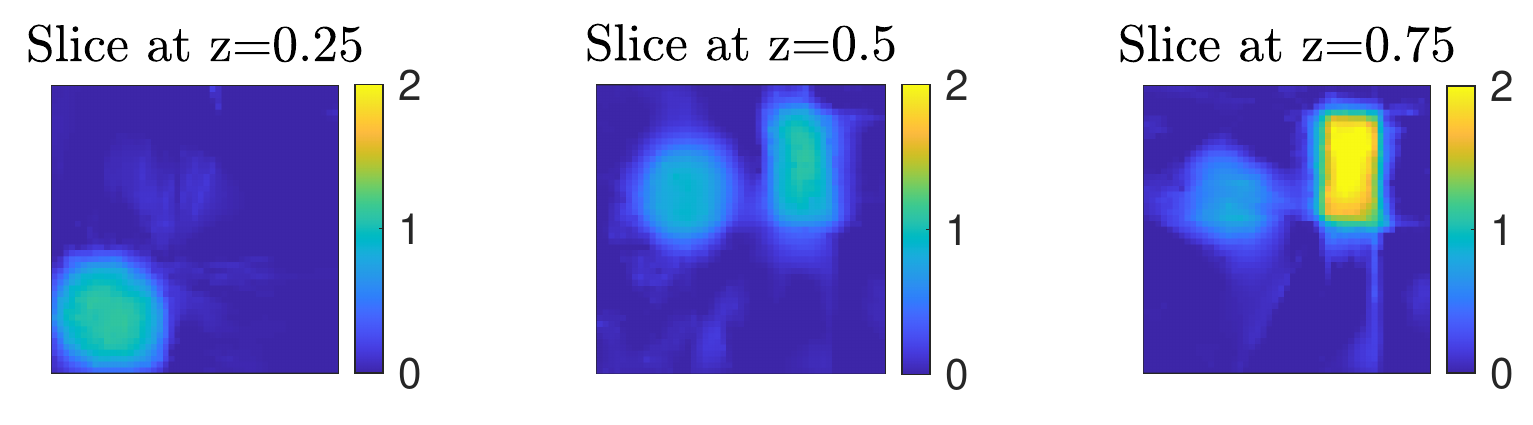}
	\caption{{\sc 3D~Test.} Slices parallel to the xy-plane of reconstructions produced by Algorithm~\ref{alg:basic_optimization} for quarter-aperture projections. Top row: 5 optimized projections. Bottom row: 15 optimized projections.}
	\label{fig:3Dtestrec10_30a}
\end{figure}

The top row of Figure~\ref{fig:3Dtestrec10_30a} shows three slices of the reconstruction parallel to the xy-plane after 5 rounds of Algorithm~\ref{alg:basic_optimization}, while the bottom row illustrates the same cross-sections after 15 rounds. These images demonstrate that initially the optimization procedure focuses solely on the vicinity of the ball centered at $(0.2,0.2,0.2)$, while the surroundings of the two other inclusions are left unexplored. This explains the rapid drop in the relative $L^2(D)$ reconstruction error over the first couple of iterations as one of the two balls is found and explored, but it also gives a reason for the slow convergence between 3 and 6 iterations: the algorithm prefers to first thoroughly investigate the detected ball, and it moves its focus on the other two objects only after an optimized projection accidentally passes through them. This demonstrates an inherent flaw in the algorithm: areas with already detected distinguishable features are examined in depth, whereas other areas are left untouched until something interesting is found as a byproduct of the ongoing local exploration. This feature could possibly be mitigated,~e.g.,~by initializing the algorithm with a low number of full-width projections that cover the entire target.

\section{Concluding remarks}
\label{sec:conclusion}

In this work we studied sequential edge-promoting Bayesian experimental design for linear inverse problems and, in particular, for X-ray tomography. We introduced a novel greedy iterative method that aims at optimizing the measurement design when a TV type prior is applied. The method is based on interpreting the so-called lagged diffusivity iteration~\cite{Vogel96} in the Bayesian framework. Our two and three-dimensional numerical examples based on simulated data suggest that the introduced approach promotes sequential designs that enhance recovery of edges in the target image.

There are a number of interesting avenues for future work. Due to the feedback from the data, our sequential algorithm often allocates subsequent projections to enhance already observed edges while a portion of the target image may remain uninvestigated. Such choices are not necessarily globally optimal, and we recorded reconstruction error plots that exhibit occasional jumps when previously unexplored objects are (accidentally) detected. Understanding the algorithmic balance between exploring new areas and improving already observed edges via the choice of the next design seems an interesting task.

The more straightforward questions are related to the performance of the algorithm for nonlinear inverse problems and its integration with more efficient optimization procedures than the exhaustive search employed here. Moreover, investigating whether the sequential designs obtained via the proposed approach approximate (at least asymptotically) the ones corresponding to the exact TV prior is also left for future studies.

\bibliographystyle{acm}
\bibliography{adapXray-refs}
\end{document}